\documentstyle[twoside,epsf]{article}

\catcode`\@=11
\long\def\@makefntext#1{
\protect\noindent \hbox to 3.2pt {\hskip-.9pt  
$^{{\eightrm\@thefnmark}}$\hfil}#1\hfill}		

\def\thefootnote{\fnsymbol{footnote}}
\def\@makefnmark{\hbox to 0pt{$^{\@thefnmark}$\hss}}	
	
\def\ps@myheadings{\let\@mkboth\@gobbletwo
\def\@oddhead{\hbox{}
\rightmark\hfil\eightrm\thepage}   
\def\@oddfoot{}\def\@evenhead{\eightrm\thepage\hfil
\leftmark\hbox{}}\def\@evenfoot{}
\def\sectionmark##1{}\def\subsectionmark##1{}}

\oddsidemargin=\evensidemargin
\renewcommand{\thefootnote}{\fnsymbol{footnote}}

\newcounter{sectionc}\newcounter{subsectionc}\newcounter{subsubsectionc}
\renewcommand{\section}[1] {\vspace{12pt}\addtocounter{sectionc}{1} 
\setcounter{subsectionc}{0}\setcounter{subsubsectionc}{0}\noindent 
	{\tenbf\thesectionc. #1}\par\vspace{5pt}}
\renewcommand{\subsection}[1] {\vspace{12pt}\addtocounter{subsectionc}{1} 
	\setcounter{subsubsectionc}{0}\noindent 
	{\bf\thesectionc.\thesubsectionc. {\kern1pt \bfit #1}}\par\vspace{5pt}}
\renewcommand{\subsubsection}[1] {\vspace{12pt}\addtocounter{subsubsectionc}{1}
	\noindent{\tenrm\thesectionc.\thesubsectionc.\thesubsubsectionc.
	{\kern1pt \tenit #1}}\par\vspace{5pt}}
\newcommand{\nonumsection}[1] {\vspace{12pt}\noindent{\tenbf #1}
	\par\vspace{5pt}}

\newcounter{appendixc}
\newcounter{subappendixc}[appendixc]
\newcounter{subsubappendixc}[subappendixc]
\renewcommand{\thesubappendixc}{\Alph{appendixc}.\arabic{subappendixc}}
\renewcommand{\thesubsubappendixc}
	{\Alph{appendixc}.\arabic{subappendixc}.\arabic{subsubappendixc}}

\renewcommand{\appendix}[1] {\vspace{12pt}
        \refstepcounter{appendixc}
        \setcounter{figure}{0}
        \setcounter{table}{0}
        \setcounter{lemma}{0}
        \setcounter{theorem}{0}
        \setcounter{corollary}{0}
        \setcounter{definition}{0}
        \setcounter{equation}{0}
        \renewcommand{\thefigure}{\Alph{appendixc}.\arabic{figure}}
        \renewcommand{\thetable}{\Alph{appendixc}.\arabic{table}}
        \renewcommand{\theappendixc}{\Alph{appendixc}}
        \renewcommand{\thelemma}{\Alph{appendixc}.\arabic{lemma}}
        \renewcommand{\thetheorem}{\Alph{appendixc}.\arabic{theorem}}
        \renewcommand{\thedefinition}{\Alph{appendixc}.\arabic{definition}}
        \renewcommand{\thecorollary}{\Alph{appendixc}.\arabic{corollary}}
        \renewcommand{\theequation}{\Alph{appendixc}.\arabic{equation}}
        \noindent{\tenbf Appendix \theappendixc #1}\par\vspace{5pt}}
\newcommand{\subappendix}[1] {\vspace{12pt}
        \refstepcounter{subappendixc}
        \noindent{\bf Appendix \thesubappendixc. {\kern1pt \bfit #1}}
	\par\vspace{5pt}}
\newcommand{\subsubappendix}[1] {\vspace{12pt}
        \refstepcounter{subsubappendixc}
        \noindent{\rm Appendix \thesubsubappendixc. {\kern1pt \tenit #1}}
	\par\vspace{5pt}}

\newcommand{\textlineskip}{\baselineskip=13pt}
\newcommand{\smalllineskip}{\baselineskip=10pt}

\def\eightcirc{
\begin{picture}(0,0)
\put(4.4,1.8){\circle{6.5}}
\end{picture}}
\def\eightcopyright{\eightcirc\kern2.7pt\hbox{\eightrm c}} 

\newcommand{\copyrightheading}[1]
	{\vspace*{-2.5cm}\smalllineskip{\flushleft
	{\footnotesize International Journal of Modern Physics B, #1}\\
	{\footnotesize $\eightcopyright$\, World Scientific Publishing
	 Company}\\
	 }}

\newcommand{\publisher}[2]{{\begin{center}\footnotesize\smalllineskip 
	Received #1\\
	Revised #2
	\end{center}
	}}

\def\abstracts#1#2#3{{
	\centering{\begin{minipage}{4.5in}\baselineskip=10pt\footnotesize
	\parindent=0pt #1\par 
	\parindent=15pt #2\par
	\parindent=15pt #3
	\end{minipage}}\par}} 

\renewenvironment{thebibliography}[1]			
	{\frenchspacing
	 \ninerm\baselineskip=11pt
	 \begin{list}{\arabic{enumi}.}
	{\usecounter{enumi}\setlength{\parsep}{0pt}
	 \setlength{\leftmargin 12.7pt}{\rightmargin 0pt} 
	 \setlength{\itemsep}{0pt} \settowidth
	{\labelwidth}{#1.}\sloppy}}{\end{list}}

\newcounter{itemlistc}
\newcounter{romanlistc}
\newcounter{alphlistc}
\newcounter{arabiclistc}

\newcommand{\fcaption}[1]{
        \refstepcounter{figure}
        \setbox\@tempboxa = \hbox{\footnotesize Fig.~\thefigure. #1}
        \ifdim \wd\@tempboxa > 5in
           {\begin{center}
        \parbox{5in}{\footnotesize\smalllineskip Fig.~\thefigure. #1}
            \end{center}}
        \else
             {\begin{center}
             {\footnotesize Fig.~\thefigure. #1}
              \end{center}}
        \fi}

\newcommand{\tcaption}[1]{
        \refstepcounter{table}
        \setbox\@tempboxa = \hbox{\footnotesize Table~\thetable. #1}
        \ifdim \wd\@tempboxa > 5in
           {\begin{center}
        \parbox{5in}{\footnotesize\smalllineskip Table~\thetable. #1}
            \end{center}}
        \else
             {\begin{center}
             {\footnotesize Table~\thetable. #1}
              \end{center}}
        \fi}

\def\@citex[#1]#2{\if@filesw\immediate\write\@auxout
	{\string\citation{#2}}\fi
\def\@citea{}\@cite{\@for\@citeb:=#2\do
	{\@citea\def\@citea{,}\@ifundefined
	{b@\@citeb}{{\bf ?}\@warning
	{Citation `\@citeb' on page \thepage \space undefined}}
	{\csname b@\@citeb\endcsname}}}{#1}}

\newif\if@cghi
\def\cite{\@cghitrue\@ifnextchar [{\@tempswatrue
	\@citex}{\@tempswafalse\@citex[]}}
\def\citelow{\@cghifalse\@ifnextchar [{\@tempswatrue
	\@citex}{\@tempswafalse\@citex[]}}
\def\@cite#1#2{{$\null^{#1}$\if@tempswa\typeout
	{IJCGA warning: optional citation argument 
	ignored: `#2'} \fi}}

\def\pmb#1{\setbox0=\hbox{#1}
	\kern-.025em\copy0\kern-\wd0
	\kern.05em\copy0\kern-\wd0
	\kern-.025em\raise.0433em\box0}

\def\fnt#1#2{\footnotetext{\kern-.3em
	{$^{\mbox{\scriptsize #1}}$}{#2}}}

\def\fpage#1{\begingroup
\voffset=.3in
\thispagestyle{empty}\begin{table}[b]\centerline{\footnotesize #1}
	\end{table}\endgroup}

\font\tenrm=cmr10
\font\tenit=cmti10 
\font\tenbf=cmbx10
\font\bfit=cmbxti10 at 10pt
\font\ninerm=cmr9
\font\nineit=cmti9
\font\ninebf=cmbx9
\font\eightrm=cmr8

\def\qed{\hbox{${\vcenter{\vbox{			
   \hrule height 0.4pt\hbox{\vrule width 0.4pt height 6pt
   \kern5pt\vrule width 0.4pt}\hrule height 0.4pt}}}$}}

\renewcommand{\thefootnote}{\fnsymbol{footnote}}	

\def\bsc{{\sc a\kern-6.4pt\sc a\kern-6.4pt\sc a}}	
\def\bflatex{\bf L\kern-.30em\raise.3ex\hbox{\bsc}\kern-.14em 
T\kern-.1667em\lower.7ex\hbox{E}\kern-.125em X} 

\begin{document}


\normalsize\textlineskip
\thispagestyle{empty}
\setcounter{page}{1}

\copyrightheading{}			

\vspace*{0.88truein}

\fpage{1}

\def\be{\begin{equation}} \def\ee{\end{equation}} \def\bea{\begin{eqnarray}}
\def\eea{\end{eqnarray}}

\centerline{\bf QUASI-LONG RANGE ORDER IN GLASS STATES OF IMPURE}
\vspace*{0.035truein}
\centerline{\bf LIQUID CRYSTALS, MAGNETS, AND SUPERCONDUCTORS}
\vspace*{0.37truein}
\centerline{\footnotesize D.E. FELDMAN}
\vspace*{0.015truein}
\centerline{\footnotesize\it 
Condensed Matter Physics Department, Weizmann Institute of Science} 
\baselineskip=10pt
\centerline{\footnotesize\it Rehovot, 76100, Israel} 
\vspace*{10pt}
\centerline{\normalsize and}
\vspace*{0.015truein}
\centerline{\footnotesize\it Landau Institute for Theoretical Physics}
\baselineskip=10pt
\centerline{\footnotesize\it Chernogolovka, Moscow region, 142434, Russia}
\vspace*{0.225truein}
\publisher{(received date)}{(revised date)}

\vspace*{0.21truein}
\abstracts{We consider glass states of several disordered systems:
vortices in impure superconductors, amorphous
magnets, and nematic liquid crystals in random porous media.
All these systems can be described by the random-field or random-anisotropy
$O(N)$ model. Even arbitrarily weak disorder destroys long range
order in the $O(N)$ model. We demonstrate that at weak disorder and low temperatures
quasi-long range order emerges. In quasi-long-range-ordered phases
the correlation length is infinite and correlation functions obey 
power dependencies on the distance. In pure systems quasi-long range order
is possible only in the lower critical dimension and only in the case of Abelian
symmetry. In the presence of disorder this type of ordering turns out to be
more common. It exists in a range of dimensions
and is not prohibited by non-Abelian symmetries.}{}{}

\vspace*{1pt}\textlineskip
\section{Introduction}
\vspace*{1pt}\textlineskip
\noindent
Solids resist small perturbations and possess some (e.g.  crystal) order.
Liquids and gases are not ordered and offer no resistance to a static shear
stress.  An intermediate class of substances which possess ordering but respond
strongly to small external disturbances is known as soft matter.  The 
fourth and last
possibility, no ordering and weak response to weak perturbations, is represented
by glass phases of strongly disordered systems.  Where should weakly disordered
systems be put in this classification?  The answer depends on their symmetry
with respect to transformations of the order parameter.
If it is discrete, weak disorder is irrelevant and the system belongs
to the class of its pure analog  
(e.g. the ground state of the pure and weakly disordered Ising ferromagnets
is the same 'spin solid').
On the other hand,
if the symmetry group is continuous then even
arbitrarily weak disorder may lead to the formation of a glass state.  However,
recent studies suggest that in many cases glass phases of weakly disordered
systems are qualitatively different from glass states at strong disorder and can
be considered as a new type of soft condensed matter (i.e.  combine ordering and
strong response to weak disturbances).  A common feature of these novel glass
phases is quasi-long range order (QLRO).

As disorder is strong, the order parameter depends only on the local 
potential of impurities and
hence ordering is absent.  In the case of weak disorder there
is a competition between internal interactions that tend to establish long range
order 
at low temperatures
and randomness that works in the opposite direction.  Naively one could
expect that stronger internal interactions win and weak disorder has no pronounced
effect on the system.  This point of view is supported by the mean field
theory and for many years was generally accepted in the field of
amorphous magnets.  However, it is not true and even arbitrarily
weak disorder may be sufficient to destroy long ranger order.  This fact was first
understood in the context of vortex lattices in impure
superconductors\cite{L} but is valid in any system of continuous symmetry.\cite{IM}
The explanation of the effect is based on the low energy
cost of large-scale collective excitations (Goldstone modes).  In three
dimensions this energy cost turns out to be lower than the energy gain due to
interaction of the large-scale excitations with disorder.  Here the continuous
symmetry which is a prerequestive for the existence of Goldstone modes is
crucial.  In the systems of discrete symmetry weak disorder is indeed not
important.
Note that if disorder is weak the correlation length is large. This
sometimes led to a wrong interpretation of experimental results 
about weakly disordered systems as evidences of long range order.

Two questions about impure systems of continuous symmetry immediately arise:
Whether the phase transition between high- and low-temperature phases survives
in the presence of disorder, and if yes, what is the nature of the
low-temperature state?  
Recently these questions have attracted a renewed experimental and theoretical
interest. On the one hand, it was stimulated by new types of disordered systems:
superfluid Helium and liquid crystals in low-density aerogels. On the other hand,
an important progress was achieved in studies of vortex lattices in impure
superconductors after high $T_c$ superconductivity was discovered. In particular,
it was found that there are two different glass states in disordered superconductors
(Vortex glass and Bragg glass).\cite{XYreview,Natrev} The Bragg glass state is observed
at weak disorder and the properties of this phase are much closer to the pure
Abrikosov state than those of Vortex glass. It turns out that
in contrast to Vortex glass, Bragg glass possesses topological order.
Topologically ordered glass phases are predicted in some other weakly disordered
systems and are the subject of the present review.

Since there is no long-range order in impure systems of continuous symmetry, 
low-temperature phases fall into the class of glass states which are known to be
a difficult subject for theory.  A usual theoretical approach in condensed
matter physics is based on an exact solution of a simple related problem.  Then
one can develop e.g.  a perturbative expansion to take into account the
ingredients missed in the exactly solvable problem.  A standard source of simple
and useful exact solutions is mean field models.  However, for strongly
disordered systems the mean field approximation is often neither simple nor very
useful.  An example is the classical solution of the mean-field spin
glass by Parisi.\cite{Parisi} If this very nontrivial result does capture
physics of real short range spin glasses is still an open question.

Much effort was devoted to theoretical understanding of
strongly disordered magnets, liquid crystals, vortex states of
impure superconductors, and other related problems, but the lack of useful exactly
solvable models has limited progress in the field.  On the other hand, one could
hope that for weakly disordered systems an appropriate solvable model is just an
analogous system without disorder, and impurities can be taken into account with a
perturbation theory.  Apparently, the ability of weak randomness to destroy long
range order, which is present in pure systems, is an argument against such an
optimistic view.  
On a more rigorous level, the exact
solution of the spherical model of amorphous 
magnets\cite{Ginz}$^-$\cite{Boyan} 
suggests that there is no qualitative difference between weakly
and strongly disordered systems with continuous symmetry.  Only recently has it
been realized that this insensitivity to the disorder strength is an artifact
of the spherical approximation, and a similarity between ordering in weakly
disordered continuous systems and their pure analogs does exist.  Long range
order is prohibited in the systems of continuous symmetry in the presence of
impurities, but instead quasi-long range order
can emerge.
This means that the average value of the order parameter over the
volume is zero, but the correlation length is infinite and correlation
functions obey power dependencies on the distance.  In other words, long range
order is only weakly broken.

The concept of quasi-long range order (QLRO) was first introduced in the pure
two-dimensional XY model.\cite{2DXY}  Due to Abelian symmetry the degeneracy
space of the XY model (i.e.  the manifold each point of which represents a
ground state) is locally indistinguishable from the degeneracy space of the free
field.  Thus, neglecting topological defects (this is legitimate at low temperatures)
one can map the pure XY model onto the free field.  This mapping gives an easy
way to understand the low-temperature behavior of the XY model.  
The result is that long range order is present in spatial dimensions $D>2$ 
and 
absent in lower dimensions.  In two dimensions
there is an intermediate situation
of QLRO in the
low-temperature phase.  
Our experience with pure systems
shows that QLRO is a rare phenomenon.  It is possible only in
Abelian systems and only in the lower critical dimension that separates the
regime of long range order in all higher dimensions from the regime with neither
long nor quasi-long range order in all lower dimensions.

Recent theoretical studies of impure systems\cite{XYnat}$^-$\cite{Feldman2} 
show however that QLRO can be much
more common in the problems with disorder than in homogeneous systems.  First, in
a given system it can occur in a whole range of dimensions and not only at the
lower critical dimension.
Second, it is not prohibited by non-Abelian symmetry.\cite{Feldman0}$^-$\cite{Feldman2}
The possibility of QLRO was predicted in
the Abrikosov state of impure superconductors,\cite{XYreview,Natrev} in
amorphous magnets\cite{Feldman0,Feldman1} and in uniaxial nematics confined in
random porous matrices.\cite{Feldman2}
Although more work is needed to completely understand glass states of weakly
disordered systems, and in particular numerical and experimental results are
contradictory, we believe that
this is only a small portion of the
list of disordered systems which could 
possess QLRO phases.  
Note that QLRO is possible only
in weakly disordered systems.\cite{Feldman3}
The nature of glass states at strong disorder and
disorder driven phase transitions from the weak-disorder to strong-disorder
regime is an important open problem which is beyond the scope of the present
review.  A very recent discussion of this question in the context of disordered
superconductors\cite{XYreview,Natrev} can be found in Ref.\cite{Nature}
Note also that the existence of a QLRO state is sensitive to
details of the system.  For example, it exists in the random-anisotropy
Heisenberg model but is absent\cite{Feldman1} in the very similar random-field
Heisenberg model that describes relaxor ferroelectrics.\cite{rfer2}

From the technical point of view the theoretical prediction
of QLRO is based on the
renormalization group applied to the weak-disorder fixed point.  It turns out
that an infinite set of relevant operators emerges in the problem and dealing
with them requires some care.  This technical problem takes its  origin in the
complicated structure of the energy landscape that contains a plethora of energy
minima.  The latter is certainly not surprising for a glass state.  Early
attempts to apply the renormalization group to this class of problems (e.g.
Ref.\cite{DF}) failed due to an incorrect treatment of the complicated energy
landscape:  Theory\cite{DF}
incorrectly predicted the dimensional reduction by 2 in
comparison with corresponding pure systems 
and suggested that QLRO
is possible only in 4-dimensional random systems of Abelian symmetry.

In this paper we discuss recent theoretical results on the nature of glass states of weakly
disordered systems with continuous symmetry. 
 Since the problem of the Bragg glass
state of the vortex lattice in impure superconductors was recently reviewed in
excellent papers,\cite{XYreview,Natrev} we consider impure superconductors
only briefly and concentrate on 
systems of non-Abelian symmetry: disordered magnets and
nematic liquid crystals.  The
central point is that in many cases the Larkin-Imry-Ma picture\cite{L,IM} of
uncorrelated domains breaks down and quasi-long range order emerges.

The article is organized as follows.  In the next section we introduce 
models of some relevant
systems:  amorphous magnets, 
relaxor ferroelectrics, vortex lattices in disordered superconductors and
uniaxial nematics in random porous media,
and map them onto the random-field (RF) and
random-anisotropy (RA) $O(N)$ models.  Section 3 contains a qualitative
discussion of the glass phases.  In section 4 we derive the functional
renormalization group equations.  Their solution is given in section 5.
First we discuss a simple approximate solution and then proceed with a
systematic approach.  Some details of the systematic solution are summarized in
the Appendix.  In section 6 we compare the predictions with experiments and
numerical experiments.  
In that section we also describe the rich phase diagram of disordered nematic.
This phase diagram includes, in particular, two QLRO glass states.
Section 7 contains conclusions.

\textheight=7.8truein
\setcounter{footnote}{0}
\renewcommand{\thefootnote}{\alph{footnote}}

\section{Disordered Systems of Continuous Symmetry}
\noindent
In this section we introduce several disordered systems of continuous symmetry.
Albeit physically very different they all can be described by
the random-field or random-anisotropy
$O(N)$ models (or slight modifications of these models).
Note that the problems of nematics and vortex lattices in superconductors
can be mapped onto the $O(N)$ models in the low-temperature phases only.
Mapping becomes invalid near the phase transitions to the high-temperature
phases.

\subsection{Amorphous magnet}
\noindent
If a solid is obtained with a fast freezing from a liquid the atoms
may have not enough time to form a crystal lattice. In this case an amorphous 
solid is formed. One can imagine it as a liquid in which the positions
of all particles are suddenly fixed. Similar to crystals amorphous solids
have special directions in each point but these directions are different 
in different places and uncorrelated for distant points. In a magnetic system,
special directions lead to the appearance of easy magnetic axes or planes.
In contrast to crystals, in amorphous magnets the directions of easy axes
are random. The simplest model
of amorphous ferromagnets is hence the random-anisotropy (RA) Heisenberg model.\cite{HPZ}
It has the following Hamiltonian:

\be 
\label{1n}
H=-J\sum_{\langle ij \rangle} {\bf n}_i {\bf n}_j - D\sum_i ({\bf n}_i{\bf h}_i)^2,
\ee
where $J$ is the exchange strength, $D$ the anisotropy strength, 
${\bf n}_i$ the spin in site $i$, $\langle ij \rangle$ denotes
a pair of neighboring sites,
and the unit vector ${\bf h}_i$ describing
the direction of random anisotropy in site $i$ varies from site to site.
 
At low temperatures amorphous magnets possess a spin-glass-like state.
This fact can be easily understood
as random anisotropy is strong 
since in this case the spins are frozen along
their local anisotropy axis. 
A review on strongly disordered amorphous magnets can be found in Ref.\cite{OS} 
In this paper we concentrate on the case of weak anisotropy.
We shall see that in the latter case there are strong correlations
between spins and the system possesses QLRO. An important consequence
of this prediction is divergence of the magnetic susceptibility
in the low-temperature phase.

\subsection{Nematic in aerogel}
\noindent
Quenched disorder is inevitably present even in the most pure solids.  This
explains a lot of phenomena, e.g.  the residual resistance of metals.  On the
other hand, liquids are usually homogeneous and introducing quenched disorder in
them requires special efforts.  One of the approaches consists in pouring a
liquid into a randomly interconnected network of pores.  Such
liquid-porous-matrix systems emerge in many natural and technological processes
giving rise to a lasting scientific activity.  The recent surge of interest in
the field is due to a new micropore material:  
low-density silica aerogel.\cite{Aerogel}
Its density can
be varied in a wide range up to more than 99\% void volume fraction.  This
allows the investigation of both strongly and weakly confined fluids.  The most
interesting situation emerges in systems with many degrees of freedom, e.g.
He-3 \cite{He3} and liquid crystals.\cite{Wu92}$^-$\cite{Bel2000}
In
these substances the porous matrix not only geometrically confines the liquid
but also induces a random orienting field that fixes the direction of the order
parameter near the surface of the matrix.
The random-field disorder is
known to cause spin-glass effects\cite{RN} and such phenomena were indeed
observed experimentally in liquid-crystal-aerogel systems.  In particular, a
slow glassy dynamics was reported in Refs.\cite{Wu92,ammy}  

Our aim is to describe the glass state of nematics in random porous media
in the case of weak disorder. This corresponds e.g. to a weak interaction with
the surface of a porous matrix. We postpone the discussion 
of the disorder strength to section 6 and formulate the simplest model
of disordered uniaxial nematics. This is a slight complification of the
RA Heisenberg model (\ref{1n}).
The free energy density 

\be
\label{2n}
F=F_{\rm d}+F_{\rm pm}
\ee
of the nematic in the porous
matrix includes the Frank distortion energy\cite{dG} 

\be
\label{3n}
F_{\rm d}=[K_1({\rm
div}{\bf n})^2+K_2({\bf n}{\rm curl}{\bf n})^2+K_3({\bf n}\times{\rm curl}{\bf
n})^2]/2, 
\ee
where ${\bf n}$ is the director, and the interaction $F_{\rm pm}$
with the surface of the random matrix.  The interaction tends to align the
director parallel to the surface.\cite{star}  We model the interaction as

\be
\label{4n}
F_{\rm pm}=({\bf hn})^2, 
\ee
where ${\bf h}$ is a random vector representing the
normal to the surface.  This is the simplest choice compatible with the
equivalence of the opposite orientations of the director.  Due to the
universality the model captures all large-scale physics. 
The average amplitude
of the random vector ${\bf h}$ is a measure of the disorder strength.  It is a
phenomenological parameter which depends on the pore size, anchoring energy and
fractal structure of the porous matrix.

We demonstrate that the model (\ref{2n}-\ref{4n}) possesses QLRO 
in its low-temperature phase at weak disorder. 
As a consequence the light-scattering 
cross-section diverges at small angles.
We also consider effects
of external electric and magnetic fields, and mechanical deformations of
the porous matrix. It turns out that 
at weak disorder
there are several phases
including two glass states.

\subsection{Relaxor ferroelectric}
\noindent
Relaxor ferroelectrics are interesting because of their diffuse phase transitions
extending over a finite range of temperatures. 
Most relaxor ferroelectrics are disordered ionic structures,
in particular, solid solutions.
In the best known relaxor ferroelectric\cite{rfer1}
PbMg$ _{1/3}$Nb$ _{2/3}$O$ _3$
the spontaneous polarization vanishes at all temperatures\cite{rfer3} and the
phase transition is believed to be destroyed by the randomness.\cite{rfer2}
The randomness can be described as quenched random electric fields.\cite{rfer2}
This leads to the random-field Heisenberg model of the relaxor ferroelectric.
Its Hamiltonian reads

\be 
\label{5n}
H=-\sum J_{ij} {\bf n}_i {\bf n}_j - \sum_i ({\bf n}_i{\bf h}_i),
\ee
where ${\bf n}_i$ is the local dipole moment and ${\bf h}_i$ the random
electric field. This model is certainly a simplification.
In particular, it misses nonrandom anisotropy (which is cubic
in PbMg$ _{1/3}$Nb$ _{2/3}$O$ _3$). 
Nonrandom anisotropy 
leads to a discrete symmetry group and hence
can stabilize long range (ferroelectric) order
which is destroyed by  random fields.\cite{IM}
However, it is expected\cite{rfer2} that
in PbMg$ _{1/3}$Nb$ _{2/3}$O$ _3$ the anisotropy is weak.
We demonstrate that the model (\ref{5n}) with the interaction of
the nearest neighbors only does not possess QLRO. The effect
of the long range dipole interactions which are present 
in relaxor ferroelectrics
is an interesting open question.

\subsection{Vortices in disordered superconductor}
\noindent
Disordered superconductors were recently considered 
in excellent reviews\cite{XYreview,Natrev} 
so in this subsection we only provide a basic background 
and briefly discuss mapping onto the random-field XY model. Our discussion
of that model in the subsequent sections is also brief. More details can
be found in the abovementioned reviews.

In the absence of disorder,
vortices in type II superconductors form a lattice. 
Small fluctuations of the vortices about their equilibrium positions
can be described with elastic theory. Let us denote 
by $R_i$ the equilibrium position
of the vortex in lattice site $i$. The vortex displacement 
${\bf u}(R_i,z)\equiv {\bf u}^i(z)$ relative to its equilibrium position is
a two-component vector. 
In the absence of dislocations we can assume that the displacement field
is slowly varying on the scale of the lattice spacing $a$. 
We will see in section 5 that topological defects are indeed irrelevant
at weak disorder and low temperatures.
Hence, a continuous
description can be used. 
The continuous elastic Hamiltonian has the following form

\be
\label{6n}
H_{\rm el}=\frac{1}{2}\sum_{\alpha,\beta}\int \frac{d^3 q}{(2\pi)^3}
\Phi_{\alpha\beta}(q)u_\alpha(q)u_\beta(-q),
\ee
where $\alpha,\beta=x,y$ label the coordinates, and the structure of the matrix
$\Phi_{\alpha\beta}$ for the triangular Abrikosov lattice can be found
in Ref.\cite{HTSC} For simplicity we consider below
a one-component field $u$. This corresponds to an anisotropic
superconductor such that the vortices can be displaced in one direction only.
It turns out that this simple model allows to obtain a good quantitative
description of some aspects of the large-scale behavior of
disordered isotropic superconductors. A discussion of
a more accurate model taking into account two components of the displacement
field and the triangular symmetry of the Abrikosov lattice
can be found in Refs.\cite{Natanis,Natanis1} We briefly discuss the results
of Refs.\cite{Natanis,Natanis1} in section 6.
The following elastic Hamiltonian describes the one-component displacement
field:

\be
\label{7n}
H_{\rm el}=\frac{c}{2}\int {d^3 x}(\nabla u)^2.
\ee
This is the Hamiltonian of the XY model without dislocations.

The next step is to include disorder.
The pinning energy reads\cite{HTSC}

\be
\label{8n}
H_{\rm pin}=\int d^3 x \rho({\bf x}) V({\bf x}),
\ee
where the vortex density

\be
\label{8an}
\rho({\bf x})=\sum_i\delta({\bf x}-{\bf R}_i - {\bf u}_i),
\ee
and $V$ is the random pinning potential with the following statistics:

\be
\label{9n}
\langle V({\bf x}) \rangle = 0;
\langle V({\bf x})V({\bf y}) \rangle = \Delta \delta({\bf x}-{\bf y}).
\ee

Mapping on the random field XY model is little bit tricky.
We rewrite the vortex density in the following way

\bea
\label{10n}
\rho({\bf x})=\sum_i\int d^3 x'  \delta({\bf x}-{\bf x}' - {\bf u}_i({\bf x}')) 
\delta({\bf R}_i - {\bf x'}) \nonumber\\
=\rho_0 \int d^3 x'  \delta({\bf x}-{\bf x}' - {\bf u}_i) \sum_{\bf Q} \exp(i{\bf Qx}') \nonumber\\
=\rho_0 {\rm det}^{-1}[\delta_{\alpha\beta}+\partial_\alpha u_\beta]\sum_{\bf Q} \exp(i{\bf Q}[{\bf x}-{\bf u}])\nonumber\\
\sim \rho_0 [1-\partial_\alpha u_\alpha] \sum_{\bf Q} \exp(i{\bf Q}[{\bf x}-{\bf u}]),
\eea
where $\rho_0$ is the average vortex density, ${\bf Q}$ are reciprocal lattice vectors.
Substituting (\ref{10n}) into (\ref{8n}) we get
an infinite set of random contributions to the energy
of the form  
$\rho_0 [1-\partial_\alpha  u_\alpha] V_{\bf Q}({\bf x}) \exp(-i{\bf Q  u}({\bf x}))$,
where $V_{\bf Q}$ are random fields.
We are interested in the large distance behavior and hence 
keep only the most relevant terms which should be taken into account by the renormalization group.
Thus, we neglect derivatives of $u$. 
We shall see that an infinite set of other random contributions is relevant in the RG sense.
Here we only write the minimal Hamiltonian of the correct symmetry for a one-component field $u$:

\be
\label{11n}
H=\frac{c}{2}\int {d^3 x}(\nabla u)^2 + \int d^3 x h({\bf x})\cos(u({\bf x}) - \alpha({\bf x})),
\ee
where $h$ and $\alpha$ are random.
This is the Hamiltonian of the random-field XY model.

It turns out that a QLRO state (Bragg glass) emerges in impure superconductors at weak disorder
and low temperatures.
The name 'Bragg glass' is due to sharp Bragg peaks which are observed in this state because
long range order is only weakly broken.
Note that the random field $h$ depends on the vortex density $\rho_0$ and hence on the applied
magnetic field along which the vortices are directed. Since the density is inversely proportional
to the magnetic field, the effective disorder strength is small at weak magnetic fields.
Indeed, in a given sample (i.e. at a fixed pinning potential $V({\bf x}))$ the Bragg glass state is
observed at low magnetic fields.

\section{Glass States}
\noindent
In this section we qualitatively discuss glass states of systems with continuous symmetry.
We demonstrate that disorder, which breaks the continuous symmetry, makes long range order
impossible and discuss a possibility of QLRO. A more general question is the existence
of a phase transition to a glass state with or without QLRO. 
There is a simple argument in favor of such transition in weakly disordered systems
whose pure analogs have a first order transition between 
high- and low-temperature phases.\cite{Spivak}
Note that in homogeneous nematics the phase transition is of the first
order. The phase transition from vortex liquid to vortex solid in pure superconductors is
also believed to have first order.\cite{HTSC}

First, let us show that even at arbitrarily weak disorder
there is no long range order in disordered systems of continuous symmetry.
We employ the Imry-Ma argument.\cite{IM}
It can be used in any system in which disorder breaks the continuous symmetry,
but we restrict our discussion by the random-field and random-anisotropy $O(N)$
models which are considered in more details in the following sections.

We start from the the classical nonlinear $\sigma$-model with the Hamiltonian

\be 
\label{1} 
H=\int d^D x[J\sum_\mu\partial_\mu{\bf n}({\bf x})
\partial_\mu{\bf n}({\bf x}) + V_{\rm imp}({\bf x})], 
\ee 
where ${\bf n}({\bf x})$ 
is the unit vector of the $N$-component 
magnetization, $V_{\rm imp}({\bf x})$ the random
potential.  In the RF case it has the form

\be 
\label{2} 
V_{\rm imp}=-\sum_\alpha h_\alpha({\bf x})n_\alpha({\bf x});\quad
\alpha=1,...,N, 
\ee 
where the random field ${\bf h}({\bf x})$ has a Gaussian
distribution and $\langle h_\alpha({\bf x})h_\beta({\bf
x}')\rangle=A^2\delta({\bf x}-{\bf x}')\delta_{\alpha\beta}$. 
In the RA case
the random potential is given by the equation

\be 
\label{3} 
V_{\rm imp}=-\sum_{\alpha,\beta}\tau_{\alpha \beta}({\bf
x})n_\alpha({\bf x})n_\beta({\bf x});\quad \alpha,\beta=1,...,N, 
\ee 
where
$\tau_{\alpha\beta}({\bf x})$ is a Gaussian random variable,
$\langle\tau_{\alpha\beta}({\bf x})\tau_{\gamma\delta}({\bf
x}')\rangle=A^2\delta_{\alpha\gamma}\delta_{\beta\delta}\delta({\bf x}-{\bf
x}')$.  The random potential (\ref{3}) corresponds to the same symmetry as the
more conventional choice $V_{\rm imp}=-({\bf hn})^2$ used in section 2
but is more convenient for
the further discussion.

We assume that the temperature is low and thermal fluctuations are
negligible.  The Imry-Ma argument\cite{IM,P} suggests that in our problem
long-range order is absent at any dimension $D<4$.  
Let us assume that the magnetization changes by order 1 at scale $L$.
This costs the exchange energy of order $E_e\sim JV/L^2$, where $V$ is the volume.
It should be compared with a possible energy gain due to disorder potential
$E_d\sim AV/L^D\times L^{D/2}=AV/L^{D/2}$. One finds that at $D<4$ the exchange
energy $E_e<E_d$ as $L$ is large. Hence, it is favorable to destroy long range order.

One can estimate the Larkin
length, up to which there are strong ferromagnetic correlations, with the
following qualitative RG approach.  Let one remove fast modes and rewrite
the Hamiltonian in terms of the block spins, corresponding to the scale $L=ba$,
where $a$ is the ultraviolet cut-off, $b>1$.  Then let one make rescaling such
that the Hamiltonian would restore its initial form with new constants $A(L),
J(L)$.  Dimensional analysis provides estimations

\be 
\label{4} 
J(L)\sim b^{D-2} J(a);\quad A(L)\sim b^{D/2}A(a). 
\ee 
To estimate the
typical angle $\phi$ between neighboring block spins, one notes that the effective
field, acting on each spin, has two contributions:  the exchange contribution
and the random one.  The exchange contribution of order $J(L)$ is oriented along
the local average direction of the magnetization.  The random contribution of
order $A(L)$ may have any direction.  This allows one to write at low
temperatures that $\phi(L)\sim A(L)/J(L)$.  The Larkin length corresponds to the
condition $\phi(L)\sim 1$ and equals $L\sim (J/A)^{2/(4-D)}$ in agreement with
the Imry-Ma argument.\cite{IM}  If Eq.  (\ref{4}) were exact the Larkin length
could be interpreted as the correlation length.  However, there are two sources
of corrections to Eq.  (\ref{4}).  Both of them are relevant already at the
derivation of the RG equation for the pure 
system in $2+\epsilon$ dimensions.\cite{Pol,Pol1}  The first source is the renormalization 
due to the interaction and
the second one results from the magnetization rescaling which is necessary to
ensure the fixed length condition ${\bf n}^2=1$.  The leading corrections to Eq.
(\ref{4}) are proportional to $\phi^2 J, \phi^2 A$.  Thus, the RG equation for
the combination $(A(L)/J(L))^2$ reads

\be 
\label{6} 
\frac{d}{d \ln L}\left(\frac{A(L)}{J(L)}\right)^2=
\epsilon\left(\frac{A(L)}{J(L)}\right)^2+ c\left(\frac{A(L)}{J(L)}\right)^4,\quad
\epsilon=4-D 
\ee 
If the constant $c$ in Eq.  (\ref{6}) is positive the Larkin
length is the correlation length indeed.  But if $c<0$ the RG equation has a
fixed point, corresponding to the phase with an infinite correlation length.  As
seen below, both situations are possible, depending on the number $N$
of the magnetization
components. The case of the infinite correlation length corresponds to QLRO.
We shall see that QLRO is possible as $N<N_c$, where $N_c$ is a critical
number. This critical number is larger in the RA
model, since the fluctuations of the magnetization are stronger in the RF case.
Indeed, in the RF model the magnetization tends to be oriented along the random
field, whereas in the RA case there are two preferable local magnetization
directions so that the spins tend to lie in the same semispace.

The above scaling argument does not allow to understand if the low-temperature
phase is different from the high-temperature paramagnetic state in case
when QLRO is absent. 
The answer turns out to be positive, 
at least if disorder is weak and the system would have a first
order phase transition in the absence of disorder.\cite{Spivak}

Near a first order phase transition a pure system has two phases: a
disordered phase with a finite correlation length and 
an ordered one with an infinite correlation
length. In the presence of impurities the ordered state is broken into
a set of Imry-Ma domains of size $L\sim (J/A)^{2/(4-D)}$. This lowers the free energy
that would equal $F_1(T)$ in the pure system
by $E_1\sim -A(V/L^D) L^{D/2} \sim -VA^{4/(4-D)}$. As disorder is weak 
the correlation length $l$ of
the disordered phase is less than the Larkin length $L$ and we can estimate 
the free energy gain in the presence of disorder as 
$E_2\sim -V\chi A^2$, where $\chi$
is the susceptibility. The condition of the first order transition is that
the free energies of two phases $F_1(T)+E_1$ and $F_2(T)+E_2$ are equal. 
One finds the following shift of the phase transition temperature

\be
\label{21n}
\Delta T=\frac{E_2-E_1}{d(F_1-F_2)/dT}.
\ee
In 3 dimensions we obtain $\Delta T\sim A^2$.
This argument does not work for systems with second order transitions
since near the phase transition the correlation lengths of both phases are infinite.
It is also valid only at $D>2$. In two dimensions first order transitions
are prohibited in disordered systems: They are destroyed due to roughening
of domain walls.\cite{Aizemann,Aizemann1}

The last remark is that QLRO is not expected in strongly disordered systems.
This agrees with the structure of the phase diagram 
of impure superconductors.\cite{XYreview,Natrev}
For the RF $O(N)$ model the absence of QLRO at strong disorder
is proven rigorously.\cite{Feldman3}

\section{Functional Renormalization Group}
\noindent
In the previous section the RG equations are discussed from the qualitative
point of view.  Eq.  (\ref{6}) corresponds to the simple Migdal-Kadanoff approach of
the section 5.1.  In the present section we derive the RG equations in a systematic
way.

The one-loop RG equations for the $N$-component RF and RA models in $4+\epsilon$
dimensions were derived in Ref.\cite{DF}
Their solution allowed to describe
the phase transition from the ferromagnetic to paramagnetic state of the RF $O(N)$
model above 4 dimensions.\cite{Feldman4}
The RG equations in dimensions $D<4$ can be obtained by just
changing the sign of $\epsilon$ in the RG equations in $D>4$.

In our derivation of RG equations we follow Ref.\cite{Feldman1}
We use the method, suggested by Polyakov\cite{Pol,Pol1} for the pure system in
$2+\epsilon$ dimensions.  This method is technically simpler and closer to the
intuition than the other approaches.  A disadvantage of the method is the
difficulty of the generalization for the higher orders in $\epsilon$.  This
generalization requires the field-theoretical approach.\cite{ZJ}

The same consideration as in the XY\cite{12a} and random manifold\cite{FRG1,FRG2}
models suggests that near a zero-temperature fixed point in $4-\epsilon$
dimensions there is an infinite set of relevant operators.  Let us show that
after replica averaging\cite{Parisi}
the relevant part of the effective replica
Hamiltonian can be represented in the form

\be 
\label{7} 
H_R=\int d^D x\left [\sum_a\frac{1}{2T}\sum_\mu \partial_\mu{\bf
n}_a\partial_\mu{\bf n}_a - \sum_{ab}\frac{R({\bf n}_a{\bf n}_b)}{T^2} \right ], 
\ee
where $a,b$ are replica indices, $R(z)$ is some function, $T$ the temperature.
We ascribe to the field ${\bf n}$ the scaling dimension $0$.  We also assume
that the coefficient before the gradient term in (\ref{7}) is $1/(2T)$ at any
scale.  Then in the $(4-\epsilon)$-dimensional space the scaling dimension of
the temperature $\Delta_T=-2+O(\epsilon)$.  Any operator $A$ containing $m$
different replica indices is proportional\cite{FRG2} to $1/T^m$.  
Indeed, before replica averaging any term of the Hamiltonian
contains one replica index and the temperature in the minus first power.
If we expand the partition function in Taylor series any $m$-replica
term of the expansion contains $T^{-m}$. This property conserves after
disorder averaging. To obtain the effective replica Hamiltonian one
reexponentiates the disorder-average series. Then one easily sees that
the power of $1/T$ in any term of the replica Hamiltonian equals
the number of the replica indices in it.\cite{footfoot}
Hence, the
scaling dimension $\Delta_A$ of the operator $A$ satisfies the relation
$\Delta_A=4-n+m\Delta_T+O(\epsilon)$, where $n$ is the number of the spatial
derivatives in the operator.  The relevant operators have $\Delta_A\ge 0$.
Hence, the relevant operators cannot contain more than two different replica
indices.  A symmetry consideration shows that all  possible relevant
operators are included into Eq.  (\ref{7}).  The function $R(z)$ is arbitrary in
the RF case.  In the RA case $R(z)$ is even due to the symmetry with respect to
changing the sign of the magnetization.

The one-loop RG equations for the $N$-component model in $4-\epsilon$ dimensions
are obtained by a straightforward combination of the methods 
of Ref.\cite{FRG2} and Refs.\cite{Pol,Pol1}
We express each replica ${\bf n}^a({\bf x})$ of the
magnetization as a combination of fast fields $\phi_i^a({\bf x}), i=1,...,N-1$
and a slow field ${\bf n}'^a({\bf x})$ of the unit length.  We use the
representation

\be 
\label{dec} {\bf n}^a({\bf x})={\bf n}'^a({\bf
x})\sqrt{1-\sum_i(\phi_i^a({\bf x}))^2}+ \sum_i\phi_i^a({\bf x}){\bf e}_i^a({\bf
x}), 
\ee 
where the unit vectors ${\bf e}_i^a({\bf x})$ are perpendicular to each
other and the vector ${\bf n}'^a({\bf x})$.  The slow field ${\bf n}'^a$ can be
chosen in different ways.  For example, one can cut the system into blocks of
size $L\gg a$, where $a$ is the ultra-violet cut-off.  In the center ${\bf x}$
of a block the vector ${\bf n}'^a({\bf x})$ should be parallel to the total
magnetization of the block.  In other points the field ${\bf n}'^a$ should
be interpolated.  We assume that the fluctuations of the magnetization are weak,
that is $\langle\phi_i^2\rangle\ll 1$.  Then the fluctuations of the field ${\bf
n}^a$ are orthogonal to the vector ${\bf n}'^a$ because of the fixed length
constraint $({\bf n}^a)^2=1$.

To substitute the expression (\ref{dec}) into the Hamiltonian we have to
differentiate the vectors ${\bf e}_i^a$.  We use the following definition

\be 
\label{dife} 
\frac{\partial {\bf e}^a_i}{\partial x_{\mu}}= -c^a_{\mu i}{\bf
n}'^a+\sum_k f^a_{\mu, ik}{\bf e}_k^a.  
\ee 
It is easy to show\cite{Pol,Pol1} that
$\sum_{\mu i}(c^a_{\mu i})^2= \sum_{\mu}(\partial_{\mu}{\bf n}'^a)^2$.  With the
accuracy up to the second order in $\phi$ the replica Hamiltonian (\ref{7}) can
be represented as follows

\bea 
\label{Hphi} 
H_{R}=\int d^D x \Big[ \frac{1}{2T}\sum_{a} \{ (\partial_{\mu}{\bf
n}'^a)^2 (1-(\phi_i^a)^2)+ c^a_{\mu i}c^a_{\mu k}\phi^a_i\phi^a_k + (
\partial_{\mu} \phi_i^a - f^a_{\mu, ik} \phi^a_k)^2 \} & & \nonumber \\ -
\frac{1}{T^2}\sum_{ab} \{ R({\bf n}'^a{\bf n}'^b)+A^{ab}(\phi^a_i)^2+
B^{ab}_{ij}\phi^a_i\phi^b_j + C^{ab}_{ij}\phi_i^a\phi^b_j \} \Big], & & 
\eea 
where
the summation over the repeated indices $i,j,k,\mu$ is assumed and

\bea 
\label{coef} 
A^{ab}=-({\bf n}'^a{\bf n}'^b)R'({\bf n}'^a{\bf n}'^b);
B^{ab}_{ij}= ({\bf n}'^b{\bf e}^a_i)({\bf n}'^b{\bf e}^a_j) R''({\bf n}'^a{\bf
n}'^b); & & \nonumber \\ C^{ab}_{ij}= ({\bf e}^a_i{\bf e}_j^b)R'({\bf n}'^a{\bf
n}'^b) + ({\bf n}'^a{\bf e}^b_j)({\bf n}'^b{\bf e}^a_i) R''({\bf n}'^a{\bf
n}'^b).  & & 
\eea
In the last formula $R'$ and $R''$ denote the first and second
derivatives of the function $R(z)$.  We have omitted the terms of the first
order in $\phi$ in Eq.  (\ref{Hphi}).  These terms are proportional to the
products of the fast field $\phi$ and some slow fields.  Hence, they are
non-zero only in narrow shells of the momentum space.  One can show that their
contributions to the RG equations are negligible.

To obtain the RG equations we have to integrate out the fast variables
$\phi^a_i$.  Near a zero-temperature fixed point the Jacobian of the
transformation ${\bf n}\rightarrow ({\bf n}', \phi_i)$ can be ignored, since the
Jacobian does not depend on the temperature.  The integration measure is
determined from the condition that the field ${\bf n}'^a$ is a slow part of the
magnetization.  This condition imposes restrictions on the fields $\phi$.  The
expression (\ref{Hphi}) depends on the choice of the vectors ${\bf e}^a_i$
(\ref{dec}).  However, after integrating out the fields $\phi$ the Hamiltonian
can depend only on the slow part ${\bf n}'^a$ of the magnetization.  One can
make the calculations simpler, considering special realizations of the field
${\bf n}'^a$.  To find the renormalization of the disorder-induced term $R(z)$
(\ref{7}) we can assume that the field ${\bf n}'^a$ does not depend on the
coordinates.  The renormalization of the gradient energy can be determined,
assuming that the vectors ${\bf n}'^a({\bf x})$ depend on one spatial coordinate
only and lie in the same plane.  In both cases the vectors ${\bf e}_i^a$ can be
chosen such that in Eq.  (\ref{dife}) $f^a_{\mu, ik}=0$.  At such a choice the
integration measure can be omitted and the fields $\phi_i^a$ can be considered
as weakly interacting fields with the wave vectors from the interval
$1/a>q>1/L$.

To derive the one-loop RG equations we express the free energy via the
Hamiltonian (\ref{Hphi}).  Then we expand the exponent in the partition function
up to the second order in $(H_R-\int d^D x \sum_{\mu
i}(\partial_{\mu}\phi_i)^2/(2T))$ and integrate over $\phi^a_i$.  Finally, we
make a rescaling.  The vectors ${\bf e}^a_i$ can be excluded from the final
expressions with the relation $\sum_i({\bf ae}^a_i)({\bf be}^a_i)= ({\bf
ab})-({\bf an}'^a)({\bf bn}'^a)$.  In a zero-temperature fixed point the RG
equations are

\be 
\label{Tz} 
\frac{d\ln T}{d\ln L}= -(D-2) + 2(N-2)R'(1)+O(R^2,T); 
\ee
\begin{eqnarray} 
\label{Rz} 
0=\frac{dR(z)}{d \ln L}=\epsilon R(z) +
4(N-2)R(z)R'(1)- 2(N-1)zR'(1)R'(z) & & \nonumber \\
+2(1-z^2)R'(1)R''(z) +
(R'(z))^2(N-2+z^2) & & \nonumber \\
-2R'(z)R''(z)z(1-z^2)+ (R''(z))^2(1-z^2)^2, & & 
\end{eqnarray}
where the factor $1/(8\pi^2)$ is absorbed into $R(z)$ to simplify notations.
The RG equations become simpler after the substitution of the argument of the
function $R(z)$:  $z=\cos\phi$.  In terms of this new variable one has to find
even periodic solutions $R(\phi)$.  The period is $2\pi$ in the RF case and
$\pi$ in the RA case due to the additional symmetry of the RA model.  The
one-loop equations get the form

\be 
\label{8} 
\frac{d\ln T}{d\ln L}= -(D-2) - 2(N-2)R''(0)+O(R^2,T); \ee
\begin{eqnarray} 0=\frac{dR(\phi)}{d \ln L}=\epsilon R(\phi) + (R''(\phi))^2 -
2R''(\phi) R''(0) - & & \nonumber\\ \label{9} (N-2)\left [4R(\phi)R''(0)+2{\rm
ctg}\phi R'(\phi) R''(0) - \left(\frac{R'(\phi)}{\sin\phi}\right)^2\right] + O(R^3,T)
& & 
\end{eqnarray} 
Eq.  (\ref{8}) provides the following result for the scaling
dimension $\Delta_T$ of the temperature

\be 
\label{DT} 
\Delta_T=-2+\epsilon-2(N-2)R''(0).  
\ee

The two-spin correlation function is given in the one-loop order\cite{Pol,Pol1} by
the expression

\be 
\label{cf} 
\langle{\bf n}^a({\bf x}){\bf n}^a({\bf x}')\rangle= \langle{\bf
n}'^a({\bf x}){\bf n}'^a({\bf x}')\rangle 
\left(1-\left\langle\sum_i(\phi_i^a)^2\right\rangle\right).
\ee 
Hence, in the fixed point $\langle{\bf n}({\bf x}){\bf n}({\bf
x}')\rangle\sim|{\bf x}-{\bf x}'|^{-\eta}$, where

\be 
\label{11} \eta=-2(N-1)R''(\phi=0) 
\ee

Let us find the magnetic susceptibility in the weak uniform external field $H$.
We add to the Hamiltonian (\ref{7}) the term $-\sum_a\int d^Dx Hn^a_z/T$ (the
field is directed along the z-axis).  The renormalization of the field $H$ is
determined by the renormalization of the temperature (\ref{8}) and the field
${\bf n}$.  In the zero-loop order the renormalized magnetic field $h(L)$
depends on the scale as $h(L)=H\times(L/a)^2$.  Hence, the correlation length
$R_c\sim H^{-1/2}$.  The magnetization, averaged over a block of size $R_c$, is
oriented along the field.  The absolute value of this average magnetization is
proportional to $R_c^{-\eta/2}$.  This allows us to calculate the critical
exponent $\gamma$ of the magnetic susceptibility $\chi(H)\sim H^{-\gamma}$ in a
zero-temperature fixed point:

\be 
\label{15}
 \gamma=1+(N-1)R''(\phi=0)/2 .  
\ee

In Ref.\cite{DF} Eqs.  (\ref{Tz},\ref{Rz}) were derived with a different
method.  In that paper the critical behavior in $4+\epsilon$ dimensions was
studied by considering analytical fixed point solutions $R(z)$.  In the
Heisenberg model, analytical solutions are absent and they are unphysical for
$N\ne 3$.\cite{DF}  In $4-\epsilon$ dimensions appropriate analytical solutions
are absent for any $N$.  To demonstrate this let us differentiate Eq.
(\ref{Rz}) over $z$ at $z=1$.  For any analytical $R(z)$ we obtain the following
flow equation

\be 
\label{flow} 
\frac{dR'(z=1)}{d \ln L}=\epsilon R'(z=1) + 2(N-2)(R'(z=1))^2.
\ee 
At $N> 2$ the fixed point of this equation $R'(z=1)=-\epsilon/[2(N-2)]<0$.
It corresponds to the negative critical exponent $\eta$ (\ref{11}) and hence is
unphysical (two-spin correlation functions are limited and cannot grow up to infinity
as $R\rightarrow\infty$).  
However, we shall see that in the RA model some appropriate
non-analytical fixed points $R(z)$ appear.  In these fixed points
$R''(z=1)=\infty$.  In Ref.\cite{DF} the RG charges are the derivatives of the
function $R(z)$ at $z=1$.  Thus, in a non-analytical fixed point these charges
diverge.  In the systems with a finite number of the charges their divergence
implies the absence of a fixed point.  However, if the number of the RG charges
is infinite such a criterion does not work and is even ambiguous.  Indeed, the
set of charges can be chosen in different ways and e.g.  the coefficients of the
Taylor expansion about $z=0$ remain finite in our problem.

Nonanaliticity of the fixed point solution is related with a complicated structure
of the energy landscape.\cite{FRG2,BalentsM} This has important consequences
for dynamics. An interested reader may refer to Ref.\cite{BalentsM}

\section{Solution of the Renormalization Group Equations}
\noindent
In this section we solve the RG equations of the preceding section.
The main aim is to understand in which cases there is QLRO and when it
is absent. The RG equations are the same for the RF and RA $O(N)$
models. The difference between the models is the symmetry
of the fixed point solution. In the RF case we are looking for the solution
which is stable to arbitrary perturbations and describes the low-temperature
phase of a weakly disordered system. In the RA case one has to respect
the symmetry to changing the sign of the magnetization. As discussed
above in the RA case the function $R(z)$ must be even. The solution
should be stable only to the perturbations that do not break this property.

We obtain the following results\cite{Feldman1} : In the RF model, QLRO exists at $N=2$
only. In the RA case, QLRO is present in $4-\epsilon$ dimensions 
for $N<10$. The critical exponents describing the QLRO phases 
of the RA $O(N)$ model are given in Table 1.

The section is organized as follows:
In the first subsection we discuss a simple approximate solution of the RG
equations. In two next subsections we provide a systematic analysis
of the RF and RA models. Since our RG approach ignores topological defects
we have to check that they are irrelevant. This is done in the fourth
subsection.

\subsection{Migdal-Kadanoff renormalization group}
\noindent
This subsection contains a simple approximate version of the renormalization
group.  The results for the critical exponents of the XY and Heisenberg models
have a very good accuracy.  The value of the magnetization component number
$N_c$, at which QLRO disappears in the RF model, is probably exact.  However,
the critical number of the components in the RA model is underestimated.

\subsubsection{Random field}
\noindent
We use the following ansatz for the disorder-induced term in the Hamiltonian
(\ref{7}):  $R({\bf n}_a{\bf n}_b)=\alpha{\bf n}_a {\bf n}_b+\beta$, where
$\alpha$ and $\beta$ are constants.  This expression corresponds to the Gaussian
RF disorder (\ref{2}).  Below we ignore the generation of other
contributions to the function $R(z)$.  The missed contributions are related to
random anisotropies of different orders.  In terms of the angle variable $\phi$
(\ref{8},\ref{9})

\be 
\label{B1} 
R(\phi)=\alpha\cos\phi+\beta.  
\ee 
To ensure consistency we have
to truncate the RG equation (\ref{9}).  We substitute the ansatz (\ref{B1}) into
Eq.  (\ref{9}) and retain only the terms, proportional to $\cos\phi$ or
independent of $\phi$.  This leads to the following RG equation for the constant
$\alpha$ (\ref{B1})

\be 
\label{B2} 
\frac{d\alpha}{d\ln L}=\epsilon\alpha+2\alpha^2(N-3).  
\ee 
For
$N<3$ Eq.  (\ref{B2}) has a stable solution $\alpha=\epsilon/[2(3-N)]$. 
This solution corresponds to a QLRO state.
The critical exponent (\ref{11}) equals

\be 
\label{B3} 
\eta=\frac{(N-1)\epsilon}{(3-N)}.  
\ee 
At $N=2$ this result has
less than ten percent difference from the systematic theory.\cite{12a}  QLRO
disappears at $N=3$.  This is the same critical number which is found with a 
systematic approach.

For $N>3$ a fixed point exists in $4+\epsilon$ dimensions.  It describes the
transition between the ferromagnetic and paramagnetic phases.  In this fixed
point the critical exponent (\ref{B3}) satisfies the modified dimensional
reduction hypothesis.\cite{mdr}  However, this is an artifact
of the Migdal-Kadanoff approximation, since the correct value of the critical
exponent\cite{Feldman4} differs form Eq.  (\ref{B3}).
The detailed discussion of the critical exponents of the RF $O(N)$ model
in $4+\epsilon$ dimensions is beyond the scope of the present article.
Some details can be found in Ref.\cite{Feldman4}

\subsubsection{Random anisotropy}
\noindent
In this case we use the ansatz $R({\bf n}_a{\bf n}_b)= A({\bf n}_a{\bf
n}_b)^2+B$.  In terms of the variable $\phi$ Eqs. (\ref{8},\ref{9})
$R(\phi)=\alpha\cos2\phi+\beta$.  We again substitute our ansatz into Eq.
(\ref{9}) and retain the terms, proportional to $\cos 2\phi$, and the terms,
independent of $\phi$.  The RG equation for the constant $\alpha$ has the form

\be 
\label{B4} 
\frac{d\alpha}{d\ln L}=\epsilon\alpha+8(N-6)\alpha^2.  
\ee 
The
fixed point solution of this equation is $\alpha=\epsilon/[8(6-N)]$.  It
describes the QLRO phase at $N<6$.  At $N=3$ the function
$R(\phi)=\alpha\cos2\phi+\beta$ is just $R_{\omega=0}$ of section 5.3.3.
The critical exponent of the two-spin correlation function is given by the
following equation

\be 
\label{B5} 
\eta=\frac{\epsilon(N-1)}{6-N}.  
\ee 
At $N=2,3$ this value is
close to the results of the systematic approach (Table 1).

The approximate analysis suggests that QLRO disappears at $N=6$.
This is different from the result of the systematic approach
$N_c=10$.

\subsection{Random field} 


\subsubsection{$N=2$}
\noindent
For the RF XY model the one-loop RG equations (\ref{8},\ref{9}) can be solved
analytically.\cite{12a}   The solution is a periodic
function $R(\phi)$ with period $2\pi$.  In interval $0<\phi<2\pi$ 
the fixed point solution
$R(\phi)$ is given by the formula

\be 
\label{RFXYsoln}
 R(\phi)=\frac{\pi^4\epsilon}{9}\left[1/36-({\phi}/{2\pi})^2
\left(1-({\phi}/{2\pi})\right)^2\right].  
\ee 
This is a stable fixed point.  This
can be verified with the linearization of the flow equation (\ref{9}) for
small deviations from the fixed point.

The solution (\ref{RFXYsoln})
corresponds to QLRO with the critical
exponents 
(\ref{11},\ref{15})
$\eta=\pi^2/9\epsilon, \gamma=1-\pi^2/18\epsilon$.  In the first order
in $\epsilon$ the exponent $\eta$ equals the prefactor $C$ before the logarithm
in the correlation function\cite{12a} of the angles $\phi({\bf x})$ between the
spins ${\bf n}({\bf x})$ and some fixed direction:  $\langle(\phi({\bf
x}_1)-\phi({\bf x}_2))^2\rangle=C\ln|{\bf x}_1-{\bf x}_2|$.  We expect that this
coincidence does not extend to the higher orders.

\subsubsection{$N>2$}
\noindent
If $N\ne 2$ the RG equation (\ref{9}) is more complex.  Fortunately, at $N>3$
there is still a simple method to study the large-distance behavior.  The method
is based on the Schwartz-Soffer inequality\cite{SwSo} and shows that QLRO is
absent.

In Ref.\cite{SwSo} the inequality is proven for the Gaussian distribution of
the random field.  It can also be proved for the arbitrary RF 
distribution.\cite{Feldman1} We prove the inequality in the Appendix.

Let us demonstrate the absence of physically acceptable fixed points in the RF
case at $N>3$.  We derive some inequality for critical exponents.  Then we show
that the inequality has no solutions.  We use a rigorous inequality for the
connected and disconnected correlation functions\cite{SwSo}

\bea
&&\langle\langle{\bf n}({\bf q}){\bf n}(-{\bf q}) \rangle\rangle  = 
\langle{\bf n}_a({\bf q}){\bf n}_a(-{\bf q})\rangle \nonumber \\
\label{17} 
&& \qquad\qquad {} - \langle{\bf n}_a({\bf
q}){\bf n}_b(-{\bf q})\rangle \le {\rm const}\sqrt{\langle{\bf n}_a({\bf q}){\bf
n}_a(-{\bf q})\rangle},
\eea 
where ${\bf n}({\bf q})$ is a Fourier-component of
the magnetization, $a,b$ are replica indices.  The disconnected correlation
function is described by the critical exponent (\ref{11}).  The large-distance
behavior of the connected correlation function in a zero-temperature fixed point
can be derived from the expression

\be 
\label
{eqchi} 
\chi\sim\int\langle\langle {\bf n}({\bf 0}){\bf n}({\bf
x})\rangle\rangle d^D x 
\ee 
and the critical exponent of the susceptibility
(\ref{15}).  The integral in the right hand side of Eq.  (\ref{eqchi}) is
proportional to $R_c^{D-\eta_1}$, where $R_c$ is the correlation length in the
external field $H$, $\eta_1$ the critical exponent of the connected correlation
function.  For the calculation of the exponent $\gamma$ (\ref{15}) we used the
zero-loop expression of $R_c$ via $H$.  Now we need the one-loop accuracy.  In
this order $R_c\sim H^{-1/[2-(N-3)R''(0)]}$.  This allows us to get the
following equation for the exponent $\eta_1$

\be 
\label{e1} 
\eta_1=D-2-2R''(0).  
\ee 
In a fixed point Eq.  (\ref{17})
provides an inequality for the critical exponents of the connected and
disconnected correlation functions.\cite{SwSo}  The inequality has the form

\be 
\label{ein} 
2(2-D+\eta_1)\ge 4-D+\eta.  
\ee 
This allows us to obtain the
following relation

\be 
\label{18} 
4-D \le\frac{3-N}{N-1}\eta+o(R), 
\ee 
where $\eta$ is given by Eq.
(\ref{11}).  The two-spin correlation function cannot increase up to 
infinity as the distance increases.  Hence, the critical exponent $\eta$ is
positive.  At $N>3$ this is incompatible with Eq.  (\ref{18}) at small
$\epsilon$.  Thus, there are no appropriate fixed points for $N>3$.  This
suggests the strong coupling regime with a presumably finite correlation length.
We see that QLRO disappears at $N>N_c\le 3$. Numerical analysis of the RG
equations supports $N_c<3$.

Certainly, in the RF XY model\cite{12a,12b} Eq.  (\ref{18}) is satisfied.
However, the unstable fixed points
of the RG equations\cite{12a,12b} do not satisfy
the inequality.

\subsection{Random anisotropy}
\noindent
In this subsection we investigate a possibility of QLRO in the RA $O(N)$ model.
The first subsubsection is devoted to the simplest case of the XY model.  The
second subsubsection contains an inequality for the critical exponent $\eta$.  The
derivation of the inequality is analogous to 
the derivation of Eq.  (\ref{18}).  This inequality
is applied in the next subsubsections.  The third subsubsection contains the results
for the Heisenberg model.  In the last subsubsection we consider the case $N>3$.

\subsubsection{$N=2$}
\noindent
This case is studied analogously to the RF XY model.\cite{12a}  At $N=2$ the RG
equation (\ref{9}) can be solved analytically.  Its solution is a periodic
function with period $\pi$.  In interval $0<\phi<\pi$ the fixed point solution
$R(\phi)$ is given by the formula

\be 
\label{RAXYsol} 
R(\phi)=\frac{\pi^4\epsilon}{144}\left[1/36-({\phi}/{\pi})^2
\left(1-({\phi}/{\pi})\right)^2\right].  
\ee 
It is a stable fixed point.  This
can be verified with the linearization of the flow equation (\ref{9}) for
small deviations from the fixed point.  Another proof of stability is based
on the inequality of the next subsection.

The stable fixed point corresponds to the QLRO phase at low temperatures and
weak disorder.  The critical exponents $\eta=\pi^2\epsilon/36,
\gamma=1-\pi^2\epsilon/72$.

The solution (\ref{RAXYsol}) is non-analytical at $\phi=0$, since
$R^{IV}(\phi=0)=\infty$.  Hence, the Taylor expansion over $\phi$ is absent.
However, a power expansion over $|\phi|$ exists.  We shall see below that the
same behavior at small $\phi$ conserves also at other $N$.

\subsubsection{An inequality for a critical exponent} \label{sec:V.B}
\noindent
We use the same approach as in the RF model.  Since in the RA case the random
field is conjugated with a second order polynomial of the magnetization, the
Schwartz-Soffer inequality\cite{SwSo} should be applied to correlation
functions of the field $m({\bf x})=(n_z({\bf x}))^2-1/N$, where $n_z$ denotes
one of the magnetization components, $1/N$ is subtracted to ensure the relation
$\langle m\rangle=0$.

To calculate the critical exponent $\mu$ of the disconnected correlation
function we use the representation (\ref{dec}) and obtain the relation

\be 
\label{RGm} 
\langle m^a({\bf x})m^a({\bf x}')\rangle= \langle m'^a({\bf
x})m'^a({\bf x}')\rangle\left(1-\frac{2N\sum_i
\langle(\phi_i^a)^2\rangle}{N-1}\right), 
\ee 
where $a$ is a replica index,
$m'=(n'_z)^2-1/N$ the slow part of the field $m$.  One finds $\mu=-4NR''(0)$.

The critical exponent $\mu_1$ of the connected correlation function is
determined analogously to the RF case.  We apply a weak uniform field $\tilde
H$, conjugated with the field $m$, and calculate the susceptibility $dm/d\tilde
H$ in two ways.  The result for the critical exponent is
$\mu_1=D-2-2(N+2)R''(0)$.

The Schwartz-Soffer inequality provides a relation between the exponents $\mu$
and $\mu_1$.  It has the same structure as Eq.  (\ref{ein}).  Finally, we obtain
the following equation

\be 
\label{RAineq} 
\eta\ge\frac{4-D}{4}(N-1)+o(R).  
\ee 
In terms of the RG
charge $R(\phi)$ this inequality can be rewritten in the form 
\be 
\label{Rin}
R''(0)\le-\epsilon/8+o(R).  
\ee

As discussed in Ref.\cite{Feldman1} any solution of the RG equations that does not
satisfy Eq. (\ref{Rin}) describes an unstable fixed point.

\subsubsection{$N=3$} \label{sec:V.C}
\noindent
In this case we solve Eq.  (\ref{9}) numerically.  Since coefficients of Eq.
(\ref{9}) are large as $\phi\rightarrow 0$, it is convenient to use a series
expansion of the fixed-point solution $R(\phi)$ at small $\phi$.  At larger
$\phi$ the equation can be integrated with the Runge-Kutta method.  The
following expansion over $t=\sqrt{(1-z)/2}=|\sin(\phi/2)|$ holds

\bea 
\label{smphi} 
R(\phi)/\epsilon
=\frac{(N-1)a^2}{1-4(N-2)a}+2a\sin^2\frac{\phi}{2}
\pm\frac{4\sqrt{2}}{3}\sqrt{\frac{-a+2(N-2)a^2}{N+2}}|\sin^3\frac{\phi}{2}| & &
\nonumber\\
 +\left(\frac{2a}{3}-\frac{2}{3(N+4)}\right)\sin^4\frac{\phi}{2}+
O(|\sin^5\frac{\phi}{2}|), & & 
\eea 
where $a=R''(\phi=0)/\epsilon$.  We see that
the RG charge $R(\phi)$ is non-analytical at small $\phi$.  Similar to the
random manifold\cite{FRG1,FRG2} and random-field XY\cite{12a} models $R^{IV}(0)=
\infty$.

Numerical calculations show that at any $N$ the solutions, compatible with the
inequality (\ref{Rin}), have sign "$+$" before the third term of Eq.
(\ref{smphi}).  The solutions to be found are even periodical functions with
period $\pi$.  Hence, their derivative is zero at $\phi=\pi/2$.  At $N=3$ there
is only one solution that satisfies Eq.  (\ref{Rin}).  It corresponds to

\be
\label{Rpp}
R''(\phi=0)=-0.1543\epsilon.  
\ee
If this solution is stable Eqs.
(\ref{11},\ref{15}) provide the following results for the critical exponents

\be 
\label{crexp} 
\eta=0.62\epsilon; \gamma=1-0.15\epsilon.  
\ee 
All the other
solutions of Eq.  (\ref{9}) do not satisfy Eq.  (\ref{Rin}) and hence are
unstable.

We have still to test the stability of the solution found.  For this aim we use
an approximate method.  First, we find an approximate analytical solution of Eq.
(\ref{9}).  We rewrite Eq.  (\ref{9}), substituting $\omega(R''(\phi))^2$ for
$(R''(\phi))^2$.  The case of interest is $\omega=1$ but at $\omega=0$ the
equation can be solved exactly.  The solution at $\omega=1$ can then be found
with the perturbation theory over $\omega$.  The exact solution at $\omega=0$ is
$R_{\omega=0}(\phi)=\epsilon(\cos 2\phi/24+1/120)$.  The corrections of order
$\omega^k$ are trigonometric polynomials of order $2(k+1)$.  The first
correction is

\be 
\label{fc} 
R_1(\phi)=-\frac{2\omega\epsilon}{99}\cos 2\phi
+\frac{\omega\epsilon}{264}\cos 4\phi +{\rm const} 
\ee 
After the calculation of
the corrections we can write an asymptotic series for the critical exponent
$\eta$ (\ref{11}):  $\eta=\epsilon(0.67-0.08\omega+0.14\omega^2-\dots)$.  The
resulting estimation $\eta=\epsilon(0.67\pm0.08)$ agrees with the numerical
result (\ref{crexp}) well.  This allows us to expect that the stability analysis
of the solution $R_{\omega=0}$ of the equation with $\omega=0$ provides
information about stability of the solution of Eq.  (\ref{9}).

To study stability of the exact solution of the equation with $\omega=0$ is
a simple problem.  We introduce a small deviation $r(\phi)$:
$R(\phi)=R_{\omega=0}(\phi)+r(\phi)$ and write the flow equation for this
deviation:

\be 
\label{fl} 
\frac{dr(\phi)}{d\ln L}=(5r(\phi)+r''(\phi)+
r''(0)\cos2\phi)/3+{\rm const}\times r''(0).  
\ee 
It is convenient to use the
Fourier expansion $r(\phi)= \sum_m a_m\cos2m\phi$.  The flow equations for the
Fourier harmonics can be easily integrated.  We see that $a_m\rightarrow 0$ as
$L\rightarrow\infty$ for any $m>0$.  The solution is unstable with respect to
the constant shift $a_0$.  However, this instability has no interest for us,
since correlation functions do not change at such shifts.\cite{FRG2}
Indeed, a constant shift corresponds to the addition of just a random term,
independent of the magnetization, to the Hamiltonian (\ref{1}).  Thus, the RG
equation possesses a stable fixed point.  This fixed point describes the QLRO
phase with the critical exponents (\ref{crexp}).

As usual in critical phenomena, in $4$ dimensions the one-loop RG equations
allow one to obtain the exact large-distance asymptotics of the correlation
function.  In the 4-dimensional case $R(\phi)=\tilde R(\phi)/\ln L$, where
$\tilde R(\phi)$ satisfies Eq.  (\ref{9}) at $\epsilon=1$.  We obtain the
following result for the two-spin correlation function with Eq.  (\ref{cf})

\be 
\label{elg} 
\langle{\bf n}({\bf x}){\bf n}({\bf
x}')\rangle\sim\ln^{-0.62}|{\bf x}-{\bf x}'|.  
\ee

\subsubsection{$N>3$}
\noindent
Numerical analysis of Eq.  (\ref{9}) shows that solutions, compatible with Eq.
(\ref{Rin}), are absent at $N\ge 10$.  Hence, QLRO is absent for any $N\ge 10$.
This agrees with the previous results 
for the spherical model.\cite{P,Ginz}$^-$\cite{Boyan}
For each integer $N<10$ the RG equation (\ref{9}) has exactly
one solution satisfying the inequality (47).  These
solutions are described in Table 1.  In the table, $\eta$ is the critical
exponent of the two-spin correlation function, $\Delta_T$ the scaling dimension
of the temperature (\ref{DT}).

Unfortunately, it is not clear if the fixed points, found at $N>3$, survive in 3
dimensions.  A zero-temperature fixed point can exist only if the scaling
dimension of the temperature is negative.  Table \ref{table1} shows that scaling
dimension is positive in the one-loop approximation at $\epsilon=1$ and $N\ge
5$.  In the 3-dimensional $O(4)$ model the one-loop correction to the scaling
dimension $-2(N-2)R''(0)\approx 0.7\epsilon$ is close to the zero-loop
approximation $-2+\epsilon$.  Thus, the next orders of the perturbation theory
are crucial to understand what happens in 3 dimensions.

In the $O(2)$ model the scaling dimension $\Delta_T=-2+\epsilon$ 
is exact.\cite{12a,FRG2}  Hence, QLRO disappears in 2 dimensions.  In systems with
larger numbers of magnetization components fluctuations become stronger.  Thus,
one expects the absence of QLRO in all the two-dimensional $O(N)$ models.

At zero temperature Eq.  (\ref{9}) is valid independently of the scaling
dimension $\Delta_T$.  It is tempting to assume that at zero temperature,
QLRO still exists in the RA $O(N>3)$ models below the critical dimension, in
which $\Delta_T=0$.  However, the experience of the two-dimensional RF XY model
does not support such an expectation.  Recent numerical simulations show that
QLRO is absent even in the ground state of that model.\cite{2DGSsim}

\begin{table}[htbp]
\tcaption{Critical exponents of the RA $O(N)$ model.}
\label{table1} 
\centerline{\footnotesize\smalllineskip
\begin{tabular}{l c c c c c c c c} 
\hline
$N$ & 2 & 3 & 4 & 5 & 6 & 7 & 8 & 9 \\
\hline
$\eta$ & $\pi^2\epsilon/36$ & $0.62\epsilon$ & $1.1\epsilon$ &
$1.7\epsilon$ & $2.7\epsilon$ & $4.6\epsilon$ & $9.0\epsilon$ & $33\epsilon$ \\
$\Delta_T$ & $-2+\epsilon$ & $-2+1.3\epsilon$ & $-2+1.7\epsilon$ &
$-2+2.3\epsilon$ & $-2+3.2\epsilon$ & $-2+4.8\epsilon$ & $-2+8.7\epsilon$ &
$-2+30\epsilon$ \\
\hline\\
\end{tabular} }
\end{table}

\subsection{Topological defects}
\noindent
Our RG procedure is based on the decomposition (\ref{dec}) which makes sense only
if the magnetization change is slow at the microscopic scale $a$.  This condition is
broken in the core of a topological defect.  
The nature of defects depends on the particular model but we consider
different types of defects
in the same way.\cite{Feldman2} In the XY model the defects are dislocation loops.
In the Heisenberg model they are point defects. Two types of defects
are possible in nematic liquid crystals:
disclination loops and point defects.\cite{dG}  A disclination loop is a line
the rotation by $2\pi$ around which reverses the director:  ${\bf
n}\rightarrow-{\bf n}$.  The structure of a point defect is analogous to the
structure of a hedgehog in the Heisenberg model.  

Topological defects are
irrelevant at small $\epsilon=4-D$ for weak disorder.  This can be understood
from the consideration of the contribution of disclination
(dislocation) loops and
pairs of point defects of size $l\gg a$ to the RG equations at the scale $l$.
After averaging over the small-scale fluctuations the size $l$ of
topological excitations plays the role of the ultra-violet cut-off.  The
renormalized temperature is small:  $T(l)\ll 1$.  Hence, thermal
fluctuations are irrelevant.  
In the random-field problem the disorder-induced term $R({\bf n}_a{\bf
n}_b)\sim\epsilon$ in the renormalized replica Hamiltonian (\ref{7}) is of order
$\langle h^2\rangle$, where the random vector ${\bf h}$ describes the
(renormalized) random field $E_{\rm dis}=-{\bf hn}$, 
$\langle ...  \rangle$
denotes the average over the realizations of disorder.  
In the RA
problem
the disorder-induced term $R({\bf n}_a{\bf
n}_b)\sim\epsilon$ is of order
$\langle h^4\rangle$, where the random vector ${\bf h}$ describes the
(renormalized) random anisotropy $E_{\rm dis}=({\bf hn})^2$.
Inside a defect the
director change is of order $1$ at the cut-off scale.  Hence, the elastic
excitation energy determined by the renormalized Hamiltonian $H(l)$ (\ref{1})
can be compensated by the interaction with disorder only in the positions
where $h\sim 1$.  The concentration of such positions is exponentially small
$\sim\exp(-1/\epsilon)$.  Thus, defects produce corrections of order
$\exp(-1/\epsilon)$ to the RG equations and do not modify the results of the
paper qualitatively.  The concentration of the topological excitations of size
$l$ is not more than of order $l^{-D}\exp(-1/\epsilon)$.  The above discussion
is valid, if disorder is weak.  In the case of strong disorder,
topological defects are present at the microscopic scale $a$ and QLRO is absent.
Thus, topological defects may drive the system into another glass state in which
the orientation of the director is determined only by the local random
potential.  The critical strength of disorder at which QLRO disappears can be
estimated by comparison of the elastic and random contributions to the energy.

The irrelevance of topological
defects for weak disorder can also be demonstrated 
with the energy argument\cite{12b} 
modified to take into account the scale dependence of the
interaction.  A possible fractal structure of large dislocation loops can
lead to their strong suppression.\cite{KNH}

Recently the role of defects in the RF XY model was a subject of intensive
discussions.\cite{12b,KNH}$^-$\cite{vortex7}
For a review see Ref.\cite{Natrev}

\section{Comparison with Experiments and Numerical Experiments}
\noindent
In this section we discuss experimental and numerical results about QLRO in disordered
superconductors, amorphous magnets and nematics in random porous media.
In the first subsection we consider superconductors.
Our discussion is very brief. More details can 
be found in Refs.\cite{XYreview,Natrev} The second subsection is devoted
to amorphous magnets. In the last subsection we consider nematics.
Since nematics are described by the model (\ref{2n}-\ref{4n}) which differs from the
random-anisotropy Heisenberg model, we have to derive RG equations for all
three Frank constants $K_i$. It turns out that at large scales the difference from
the RA Heisenberg model is irrelevant and QLRO phase exists at low temperatures
and weak disorder. We then make predictions about light scattering in the QLRO state
and discuss several new phases that emerge in the presence of external
electric or magnetic fields, or mechanical stresses. At the end of the subsection
we consider existing experimental and numerical data concerning QLRO
in disordered nematics.

\subsection{ Disordered superconductors}
\noindent
There is a huge amount of literature about vortex states of disordered superconductors.
For example, in the latest issue of Nature before this review was submitted two important
experimental observations were published: inverse melting of the vortex 
lattice in BSCCO\cite{Nature} and discovery of 
a second vortex liquid phase in YBCO.\cite{nature2}
A particular question of QLRO in disordered superconductors has also received much attention.

The most important theoretical prediction\cite{12b}
is the emergence of Bragg peaks in scattering experiments with
disordered superconductors at weak applied magnetic fields.
The vortex density $\rho(x)$ is given by Eq. (\ref{8an}). 
In scattering experiments the square
of the modulus $|\rho_k|^2$ of the Fourier transform of the density
is measured.\cite{Ziman} For a perfect lattice
the Fourier transform is non-zero only for ${\bf k}$ equal to reciprocal
vectors of the lattice. For a non-ideal lattice we get

\be
\label{51n}
|\rho_{\bf k}|^2\sim\sum_{ij}\exp(i{\bf k}[{\bf R}_i-{\bf R}_j])
\exp(i{\bf k}[{\bf u}_i-{\bf u}_j]).
\ee
After disorder and thermal averaging one sees that the scattering
measures the Fourier transform of the correlation function

\be
\label{52n}
C_{\bf k}({\bf x})=\big\langle 
\exp(i{\bf k}[{\bf u}({\bf x})-{\bf u}({\bf 0})]) \big\rangle
=\big\langle 
\cos({\bf k}[{\bf u}({\bf x})-{\bf u}({\bf 0})]) \big\rangle.
\ee

In terms of the Hamiltonian (\ref{11n}) for the vortex displacements from the
regular positions in the Abrikosov lattice, the correlation function (\ref{cf}) is
$C_{\bf k}({\bf x})$ for a particular value of ${\bf k}$. 
We have seen that
the correlation function (\ref{cf}) obeys a power dependence on the distance.
The generalization of this property
for an arbitrary ${\bf k}$ is straightforward.\cite{XYreview,Natrev}
In terms of the scattering experiment this leads to the situation which is intermediate
between an ideal lattice and short-range order: There are Bragg peaks but they have finite widths.
These Bragg peaks were indeed observed\cite{Bragg1,Bragg2} in agreement with theory.

Numerical experiments also support QLRO in the RF XY model of disordered superconductors
(see Ref.\cite{numXY}).

Finally, we note that the RF XY model is a simplification. A more realistic
model that takes into account the triangular symmetry of the Abrikosov 
lattice\cite{Natanis,Natanis1} 
gives slightly different predictions. An interesting point
is nonuniversal QLRO: critical exponents are different in different points
of the phase diagram. However, they vary in a very limited range.\cite{Natanis,Natanis1}

\subsection{Amorphous magnets}
\noindent
A consequence of QLRO, predicted in the RA Heisenberg model, is divergence of
the magnetic susceptibility (\ref{15}). Earlier analysis of Arrott plots, which show
the field dependence of the magnetization, suggested that the susceptibility 
does diverge at low applied fields.\cite{AP} Later it become clear that in strongly
disordered amorphous magnets the zero-field susceptibility is actually finite
(e.g. Ref.\cite{vM}). What happens at weak disorder is a more difficult experimental
question. There is an evidence of a finite susceptibility\cite{B}
at weak random anisotropy. This could be interpreted as an experimental argument
in favor of the absence of QLRO in the RA Heisenberg model in $4-\epsilon$
dimensions at $\epsilon=1$. However, in such case the Imry-Ma argument\cite{AP1} and
the RG analysis in the spirit of section 3 
would predict the following scaling for the
susceptibility: $\chi\sim(J/A)^4$, where $J$
is the exchange strength, $A$ is the anisotropy. This scaling was not observed
and in contrast to the theoretical expectations
it turned out that  $\chi\rightarrow{\rm const}$
as $A\rightarrow 0$. The authors of Ref.\cite{B} interpreted this as an effect of dipole
forces. However, the magnetic susceptibility
of the pure Heisenberg ferromagnet is expected to be infinite even in the presence
of dipole forces.\cite{PPb} This suggests that the experimental system used in 
Ref.\cite{B}
cannot be described as a realization of the RA Heisenberg model 
even with dipole forces.
Besides, the effect of dipole forces on the existence of a QLRO state 
in the RA Heisenberg model is an open question.

From the numerical side the existence of QLRO is supported by a recent paper.\cite{num1}
However, numerical results\cite{num2} support also QLRO in the RF Heisenberg model
in contradiction with our predictions. We believe that this is a finite-size effect.
More work is needed for a better numerical understanding of the RA and RF
Heisenberg models.

\subsection{Disordered nematics}
\noindent
In the one-constant approximation $K_1=K_2=K_3$ the energy $F=F_{\rm d}+F_{\rm
pm}$ (\ref{2n}) reduces to the Hamiltonian of the RA Heisenberg model. 
Since that model possesses QLRO the same ordering is expected for
the randomly confined nematic.  However, in all nematics $K_1, K_3>K_2$ and this
could change the critical exponents of the correlation functions in the QLRO
state in comparison with the random Heisenberg model.  Below we demonstrate that
this is not the case, i.e.  the nematic in the porous matrix belongs to the
universality class of the RA Heisenberg model.\cite{Feldman2}
To get a simple idea why it
occurs we first consider a two-dimensional nematic film with the director ${\bf
n}=(n_{\rm x}, n_{\rm y}, n_{\rm z})=(\cos\phi , \sin\phi , 0)$ confined in the
plane $xy$ of the film in the absence of disorder.  The Frank energy is
$F_d=(K_1+K_3)({\bf \nabla}\phi)^2/2 +(K_3-K_1)\{\cos
2\phi[(\partial_x\phi)^2-(\partial_y\phi)^2]/2 +\sin
2\phi\partial_x\phi\partial_y\phi \}$.  The low-temperature phase of this system
possesses QLRO, only the term $(K_1+K_3)({\bf \nabla}\phi)^2/2$ being relevant
at large scales since $\langle\sin 2\phi\rangle=\langle\cos 2\phi\rangle=0$

The systematic consideration is based on the RG equations in $4-\epsilon$
dimensions.  Our method follows the line of section 4. 
All relevant operators of the
appropriate symmetry are included in the following replica Hamiltonian

\bea
\label{1npm} 
H_R=\int d^3 r \Big[
\frac{1}{2}\sum_a(\lambda_1\partial_\alpha n^a_\beta\partial_\alpha n^a_\beta+
\lambda_2\partial_\alpha n^a_\alpha\partial_\beta n^a_\beta 
& & \nonumber \\ 
+ \lambda_3
n^a_\alpha\partial_\alpha n^a_\beta n^a_\gamma\partial_\gamma n^a_\beta)
-\sum_{ab}\frac{R({\bf n}^a{\bf n}^b)}{T}\Big], & &
\eea
where $a,b$
are replica indices, $\alpha,\beta=x,y,z$ label the spatial coordinates,
$\lambda_1=K_2, \lambda_2=K_1-K_2,\lambda_3=K_3-K_2$, $T$ is the temperature,
the function $R(z)$ describes disorder, and the summation over the repeated
indices $\alpha$ and $\beta$ is assumed.  Due to the symmetry ${\bf
n}^a\leftrightarrow -{\bf n}^a$ the function $R(z)$ is even.  Below we measure
the temperature in units of $K_2$, and hence set $\lambda_1=1$.  To define the
energy in $4-\epsilon$ dimensions we add to the Hamiltonian (\ref{1npm}) the term
$\lambda_0\sum_{\alpha\beta}\partial_\alpha n^a_\beta \partial_\alpha
n^a_\beta/2$, where $\alpha$ labels the coordinates in the
$(1-\epsilon)$-dimensional subspace, $\beta=x,y,z$.  The stability 
conditions\cite{dG} $K_1, K_3>0$ lead to the inequality

\be 
\label{stability} 
\lambda_2,\lambda_3>-1.  
\ee

At each step of the RG procedure which is exactly the same as in section 4
we require that $\lambda_1=1$ is unchanged. 
We get two additional RG equations in comparison
with the RA and RF Heisenberg models. These equations describe the renormalization
of the elastic constants $\lambda_2$ and $\lambda_3$.
The RG equations in the first order in
$\epsilon=4-D$ read

$$
\frac{dT}{d\ln L}=-(D-2)T + (1-\lambda_3)C_\phi T;
$$
$$
\frac{d\lambda_2}{d\ln L}=-\lambda_2(1+\lambda_3)C_\phi; 
$$
\be
\label{4npm}
\frac{d\lambda_3}{d\ln
L}= -(3\lambda_3+\lambda_3^2 - \lambda_2)C_\phi, 
\ee
where the constant

\be
\label{6npm}
C_\phi=\frac{dR(z=1)/dz}{8\pi^2\sqrt{\lambda_0(1+\lambda_3)}}\left[1+\frac{1}{1+\lambda_2}\right]
\ee
describes the fluctuations of the small-scale fields (\ref{dec})

\be
\label{7npm}
\langle\phi_1^2\rangle=\langle\phi_2^2\rangle=C_\phi\ln(L/a).  
\ee
We
omit the RG equations for $\lambda_0$ and $R(z)$ since their structure is
irrelevant below.  Eqs.  (\ref{4npm}) 
have the only fixed point compatible with the
stability conditions (\ref{stability}).  In this fixed point
$T=\lambda_2=\lambda_3=0$ and Eq.  (\ref{1npm}) reduces to the Hamiltonian of the
RA Heisenberg model which thus describes the large-distance physics of the
randomly confined nematic.  Since that model possesses QLRO in its
low-temperature phase for weak disorder, QLRO is also possible in confined
nematics.  For strong disorder or high temperature the ordering disappears.
Thus, there are three phases:  the high-temperature isotropic phase and two
low-temperature glass phases with and without QLRO.  In both glass phases the
local orientation of the director is fixed by the random potential.  
The disorder driven transition between the glass phases is
related with topological defects.

The large-scale correlations of the director lead to strong small-angle light
scattering.  We determine its intensity in the limit of the weak optical
anisotropy, i.e.  assuming that in the dielectric tensor
$\epsilon_{\alpha\beta}=\epsilon_\perp\delta_{\alpha\beta}+\epsilon_an_\alpha
n_\beta$ the anisotropic term $\epsilon_a\ll\epsilon_\perp$.  In this case the
scattering cross-section can be found with the Born approximation.  The
scattering cross-section with the change of the wave vector by ${\bf q}$ is
given by the expression\cite{dG}

\be
\label{8npm} 
\sigma({\bf q})=|\omega^2/(4\pi
c^2)i_\alpha\epsilon_{\alpha\beta}({\bf q})f_b|^2, 
\ee
where $\omega$
is the light frequency, ${\bf i}$ and ${\bf f}$ are the unit vectors specifying
the initial and final polarizations, $\epsilon_{\alpha\beta}({\bf q})$ is the
Fourier transform of the dielectric tensor.  Hence, $\sigma({\bf q})\sim\langle
Q_{\alpha\beta}({\bf q})Q_{\alpha\beta}(-{\bf q})\rangle$, where
$Q_{\alpha\beta}=n_\alpha n_\beta-\delta_{\alpha\beta}/3$ is the order parameter
and the angular brackets denote the disorder and thermal averages.  In contrast
to the bulk nematic the scattering is caused not by the thermal fluctuations but
by the frozen configuration of the director.  The cross-section $\sigma({\bf
q})$ is proportional to the Fourier transform of the correlation function
$G({\bf r})=\langle Q_{\alpha\beta}({\bf 0})Q_{\alpha\beta}({\bf r})\rangle$.
In the QLRO state this correlator obeys a power dependence on the distance
$G(r)\sim r^{-\eta}$.  To calculate the exponent $\eta$ we decompose
$Q_{\alpha\beta}$ into small-scale and large-scale parts with Eq.  (\ref{dec}) and
average over the small-scale fluctuations with Eq.  (\ref{7npm}):

\bea
\label{9npm} 
\langle Q_{\alpha\beta}({\bf 0})Q_{\alpha\beta}({\bf
r})\rangle_\phi= \big\{n'_\alpha({\bf 0})n'_\beta({\bf
0})(1-\sum_i\langle\phi^2_i\rangle)+ & & \nonumber \\
\sum_{ij} e_\alpha^i({\bf 0})e^j_\beta({\bf
0})\langle\phi_i\phi_j\rangle- \delta_{\alpha\beta}/{3}\big\}\times
\big\{n'_\alpha({\bf r})n'_\beta({\bf r})(1-\sum_i\langle\phi^2_i\rangle) 
& & \nonumber \\
+\sum_{ij}
e_\alpha^i({\bf r})e^j_\beta({\bf r})\langle\phi_i\phi_j\rangle
-\delta_{\alpha\beta}/{3}\big\} = Q'_{\alpha\beta}({\bf 0})Q'_{\alpha\beta}({\bf
r})[1-6C_\phi\ln L/a], & & 
\eea
where $Q'_{\alpha\beta}=n'_\alpha
n'_\beta-\delta_{\alpha\beta}/3$, $\langle ...  \rangle$ denotes the average
over the fluctuations of $\phi$, and the relation
$\langle\phi_i\phi_j\rangle\sim\delta_{ij}$ which is valid in the RA Heisenberg
fixed point is used.  The constant $C_\phi=-2R''(0)=0.309\epsilon$
Eq.  (\ref{7npm}) is the
same as in the fixed point of the RA Heisenberg model 
(see Eq. (\ref{Rpp})).  The exponent
$\eta$ can be found with the iterative use of Eq.  (\ref{9npm}) at each RG step
until the scale $L=r$ is reached.  At the scale $r$ the values of the
renormalized director field ${\bf n}'$ are the same at the points ${\bf 0}$ and
${\bf r}$.  Hence, $Q'_{\alpha\beta}({\bf 0})Q'_{\alpha\beta}({\bf r})\sim 1$
and $r^{-\eta}\sim (1-6C_\phi\ln L/a)^K$, where $K=\ln (r/a)/\ln (L/a)$ is the
number of the RG steps.  Thus, $\eta=6C_\phi$.  The small-angle scattering
cross-section is given by the expression\cite{Feldman2}

\be
\label{10npm} 
\sigma({\bf q})\sim q^{-D+\eta}=q^{-4+2.9\epsilon}.
\ee

The uniaxial stress modifies the large-distance behavior.  The compression along
the $z$-axis can be described by adding to the Hamiltonian the term
$F_S=An_z^2$, where $A>0$, since the deformation tends to make the pore surfaces
parallel to the $xy$ plane and hence favors the planar configuration of the
director.  The uniaxial stretch is described by $F_S=An_z^2$ with a negative
$A$.  In both cases $A$ is proportional to the deformation.  The effect of the
electric field is analogous to the effect of the stress but the sign of the
electric energy\cite{dG} $F_e=-\epsilon_a({\bf nE})^2/8\pi$ is fixed for a
given substance.  The RG flow is unstable with respect to the perturbation $F_s$
and new regimes emerge at the scale $R=R_c$ at which the renormalized $A(R)\sim
1$.  The critical length $R_c$ can be found analogously to the correlation
length of the RA Heisenberg model in the uniform magnetic field (section 5).  At
small $A$ the result is $R_c\sim |A|^{-1/(2-2C_\phi)}=|A|^{-0.5-0.15\epsilon}$.
The stretched system is long-range-ordered at the scales $R>R_c$.  The nematic
order parameter can be calculated analogously to the magnetization of the RA
Heisenberg model in the uniform magnetic field and is given 
by the formula\cite{Feldman2}
$Q=\langle n_\alpha n_\beta-\delta_{\alpha\beta}/3\rangle\sim
R_c^{-3C_\phi}\sim|A|^{0.46\epsilon}$. 
Long range order can also be achieved by applying an
arbitrarily weak external magnetic field to the confined nematic since the
magnetic contribution to the energy\cite{dG} $F_m=-\chi_a({\bf nH})^2/2$ has
the same structure as the energy related with the uniaxial stretch.  A more
interesting situation emerges under the compression.  The director averaged over
a scale $R>R_c$ is confined in the $xy$-plane.  The system is thus described by
the RA XY model.  It possesses QLRO but the critical exponents are different
from the exponents of the Heisenberg model.\cite{Feldman2}  
Thus, at the scale $R_c$ the
cross-over from one QLRO state to another occurs.  Using the RA XY fixed point
found in section 5 and repeating the derivation of Eq.  (\ref{10npm}) one
finds the Born light-scattering cross-section for $q<1/R_c$:  $\sigma({\bf
q})\sim q^{-4+\epsilon(1+\pi^2/9)}.$ In the RA XY regime the cross-section
is anisotropic:  The small-angle scattering is suppressed, if the
incident or scattered light is polarized along the compression direction.
The phase diagram in the presence of a uniaxial deformation is
shown in Fig. 1.\cite{ffp}

\begin{figure}[htbp] 

\label{fig1} 

\vspace*{13pt}

\epsfxsize=3.2truein
\hskip 0.0truein
\epsffile{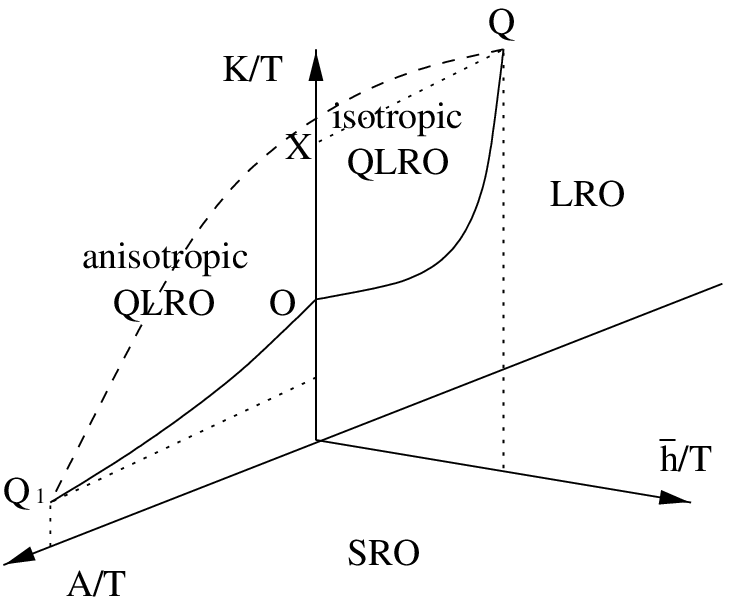}

\vspace*{13pt}

\fcaption{The phase diagram 
in the one-constant
approximation $K=K_1=K_2=K_3$.  $\bar {h}$ is the disorder strength, 
$T$ the temperature, $A$ the anisotropy.
Isotropic QLRO exists in region OXQ in the $A=0$ plane.
Surface OQQ$_1$ indicates the border between
the region of short range order (SRO) and the region with anisotropic QLRO.
At $A<0$ the system possesses long range order (LRO).}
\end{figure}

The only experimental attempt to test the existence of QLRO\cite{Bel2000}
did not provide any decisive evidence. While a power decay of correlation functions
was not observed, only distances less than the experimentally determined
Larkin length (section 3) were probed.\cite{Belpriv} However, as discussed in 
section 3,
behavior at the scales, which are less than the Larkin length, is the same
both with and without  QLRO.

Besides, the experiment\cite{Bel2000} was done with nematic confined in 
aerosil.\cite{aerosil}
This system cannot be described by the model (\ref{2n}-\ref{4n}) since disorder
is only partially quenched and elasticity-mediated nonlocal interactions can exist.
The same problem concerns a possible application of our results to nematic 
elastomers.\cite{FT,E}

One more problem is that no way to determine the disorder strength in experimental systems
was suggested. In Ref.\cite{RT} it was claimed that the effective disorder strength is less
in low-density porous media. This claim however ignores the dependence of the disorder
strength on the structure of the porous media. For example, it is evident that no ordering is possible
in the case of disconnected pores even if the density of the porous media is arbitrarily low.

Results of numerical experiments are controversial: Ref.\cite{chak} supports QLRO
but Ref.\cite{Bel2000} does not. We believe that one should take the existing
numerical results with care. In particular, there is a question about finite-size
effects in Ref.\cite{chak} while in Ref.\cite{Bel2000} disorder is not weak.
Randomness is introduced only in a small fraction of sites, but the random anisotropy
on each site is assumed to be infinitely strong. This is equivalent to the application of a 
still-not-weak random field in all six neighboring sites, and the random fields
in these six sites are
correlated. The latter also makes effective disorder stronger.
Hence, the effective disorder strength in the 
corresponding continuous model is not small.
Thus, more experimental and numerical work is needed.

\section{Conclusions}
\noindent
In conclusion, we have demonstrated that 
the random-field and random-anisotropy $O(N)$ models
possess quasi-long range order
at low temperatures and weak disorder.
In the random-field $O(N)$ model, quasi-long range order
is possible at $N=2$. In the random-anisotropy model,
QLRO exists at $N<10$.
These results can be applied to vortex states
of disordered superconductors, amorphous magnets and nematic liquid crystals
in random porous media.

There are many other related systems which we did not consider in this paper.
As the only examples we mention weakly disordered two-dimensional XY ferromagnets
with both exchange and dipole-dipole interactions\cite{Feldman5,Feldman6}
and charge density waves with long-range Coulomb interactions 
in two dimensions in the presence of disorder.\cite{Chitra}
These systems can be described as
super-quasi-long-range-ordered: The correlation
functions depend on the distance logarithmically slow similar to
the correlation function of the four-dimensional
random-anisotropy Heisenberg model (\ref{elg}).

The results about glass states of weakly disordered systems of continuous
symmetry are a part of a wider development. We have good understanding of
weakly disordered systems, if disorder breaks only the translational symmetry.
Random fields and random anisotropy also break the symmetry with respect
to transformations of the order parameter. In this case much has still to be
investigated. Even the simplest problem, random-field Ising model,\cite{RN}
is far
from being completely understood. Recent results\cite{MY,SA,Feldman4} make it
hopeful that a satisfactory theory of phase transitions in the 
random-field Ising model will be developed soon. 
Such theory should include new insights in comparison with the existing theory
of phase transitions in homogeneous systems and may open important new perspectives
in statistical mechanics in the presence of disorder.


\nonumsection{Acknowledgments}
\noindent
Useful discussions with T. Bellini, E. Domany, Y. Gefen, S.E. Korshunov,
V.V. Lebedev, V.P. Mineev,
T. Nattermann, V.L. Pokrovsky, M. Schwartz, B. Spivak
and V. Steinberg are gratefully
acknowledged. I thank R. Whitney for the critical reading of the manuscript.
 This work was supported by the Koshland fellowship and RFBR
grant 00-02-17763.

\appendix 


\noindent
In this appendix we derive an inequality for the correlation functions of the
disordered systems.  We consider a system with the Hamiltonian

\be 
\label{A1} 
H=\int dx^D [ H_1(\phi({\bf x})) - h({\bf x})m(\phi({\bf x}))],
\ee 
where $\phi$ is the order parameter, $h$ the random field with short range
correlations, $H_1$ may depend on some other random fields.  We prove the
inequality for the Fourier components of the field $m$:

\be 
\label{A2} 
G_{\rm con}({\bf q})\le {\rm const}\sqrt{G_{\rm dis}({\bf q})}, 
\ee 
where
$G_{\rm dis}({\bf q})=\overline{\langle{\bf m}({\bf q}){\bf m}(-{\bf q}) \rangle} ,
G_{\rm con}({\bf q})= \overline{\langle{\bf m}({\bf q}){\bf m}(-{\bf q})\rangle} -
\overline{\langle{\bf m}({\bf q})\rangle\langle{\bf m}(-{\bf q})\rangle}$, the
angular brackets denote the thermal averaging, the bar denotes the disorder
averaging.

Let $P(h)$ be the distribution function of the field $h$. 
Then

\bea 
G_{\rm con}({\bf q})= \int \left( P(h) \frac{d}{d h({\bf q})} m_{\bf
q}(h)\right) D\{h\}= & & \nonumber\\ 
\label{A3} 
-\int P(h) \left(\frac{d\ln P(h)}{dh({\bf
q})} m_{\bf q}(h)\right) D\{h\}, 
\eea 
where $\int D\{h\}$ denotes the integration over the
realizations of the random field, $m_{\bf q}(h)$ $=$ ${\int
D\{\phi\}\exp(-{H}/{T})m({\bf q})}$ $/$ ${\int D\{\phi\}\exp(-{H}/{T})}$.
Applying the Cauchy-Bunyakovsky inequality to Eq.  (\ref{A3}) one gets Eq.
(\ref{A2}) where the ${\rm const}=
{\rm max}_{\bf q}\sqrt{\int D\{h\} |dP(h)/dh({\bf q})|^2/P(h)}$.

For systems in the critical domain there is a simple way to understand why the
inequality is valid not only in the Gaussian case but also in a general
situation.  This is just a consequence of the universality.

\nonumsection{References}




\begin{thebibliography}{99}
\bibitem{star} Other types of anchoring do not modify the results of the paper.
\bibitem{footfoot}
The analysis of the renormalization group expansion 
for the operators $A_m\sim T^{-m}$ (Ref.\cite{FRG2}) shows
that there is no feedback proportional to $T^{-k}$ 
at $k>m$.
\end{thebibliography}
\end{document}